\documentclass[traditabstract]{aa} 

\usepackage{txfonts}
\usepackage{natbib}                             
\usepackage{graphicx}
\usepackage{lscape}
\usepackage{url}
\usepackage{longtable}

\usepackage{color}

\begin{document}

   \title{The 3XMM/SDSS Stripe 82 Galaxy Cluster Survey: }
      \subtitle{Cluster catalogue and discovery of two merging cluster candidates}


\authorrunning{Takey et al. 2016} 
\titlerunning{The 3XMM/SDSS Stripe 82 Galaxy Cluster Survey. I. }

   \author{A. Takey \inst{1,2}, 
           F. Durret\inst{1},
           E. A. Mahmoud\inst{2}, \and
           G. B. Ali\inst{2}
          }

   \institute{Sorbonne Universit\'es, UPMC Univ. Paris 6 et CNRS, UMR~7095, 
              Institut d'Astrophysique de Paris (IAP),
              98bis Bd Arago, 75014 Paris, France\\
              \email{takey@iap.fr}
         \and
             National Research Institute of Astronomy and Geophysics (NRIAG), 
             11421 Helwan, Cairo, Egypt
             }

   \date{Received ....; accepted ....}


   \abstract { We present a galaxy cluster survey based on XMM-Newton
     observations that are located in Stripe 82 of the Sloan
     Digital Sky Survey (SDSS). The survey covers an area of 11.25
     deg$^2$. The X-ray cluster candidates were selected as
     serendipitously extended detected sources from the third
     XMM-Newton serendipitous source catalogue (3XMM-DR5). A
     cross-correlation of the candidate list that comprises 94 objects
     with recently published X-ray and optically selected cluster
     catalogues provided optical confirmations and redshift estimates
     for about half of the candidate sample. We present a catalogue of
     X-ray cluster candidates previously known in X-ray and/or optical
     bands from the matched catalogues or NED. The catalogue consists
     of 54 systems with redshift measurements in the range of
     0.05-1.19 with a median of 0.36. Of these, 45 clusters have
     spectroscopic confirmations as stated in the matched
     catalogues. We spectroscopically confirmed another 6 clusters
     from the available spectroscopic redshifts in the SDSS-DR12.  The
     cluster catalogue includes 17 newly X-ray discovered clusters,
     while the remainder were detected in previous XMM-Newton and/or
     ROSAT cluster surveys.  Based on the available redshifts and
     fluxes given in the 3XMM-DR5 catalogue, we estimated the X-ray
     luminosities and masses for the cluster sample.  We also present
     the list of the remaining X-ray cluster candidates (40 objects)
     that have no redshift information yet in the literature. Of these
     candidates, 25 sources are considered as distant cluster
     candidates beyond a redshift of 0.6. We also searched for
       galaxy cluster mergers in our cluster sample and
       found two strong candidates for newly discovered cluster
       mergers at redshifts of 0.11 and 0.26. The X-ray and optical 
       properties of these systems are presented. }

\keywords{X-rays: galaxies: clusters, surveys, catalogs, galaxies: clusters: general, galaxies: clusters: intracluster medium, galaxies: clusters: individual: Abell 0412, 3XMM J030617.3-000836, 3XMM J030633.1-000350, 3XMM J010606.7+004925 and 3XMM J010610.0+005108}

\maketitle



\section{Introduction}

As galaxy clusters are the most massive gravitationally bound
objects in the Universe, their physical properties are interesting to
analyse per se\textup{ \textup{{\it }}} \citep[e.g.][]{Sarazin88, Bahcall88, Boehringer06, 
Rosati02, Ota12, Ettori13}. But they are also of interest for cosmology,
since counting clusters adds a constraint on cosmological parameters
\citep[e.g.][]{Gioia90, Voit05, Vikhlinin09a, Allen11}. This has led to 
developing systematic searches for clusters at mm, optical, and X-ray 
wavelengths because of the multi-component nature of their observed matter.
As of very recently, galaxy clusters can be detected by their Sunyaev-Zeldovich (SZ) 
signal at mm wavelengths, which provided a sample of several hundred massive 
clusters \citep[e.g.][]{Hasselfield13, Bleem15, Planck15}.    

Tens of thousands of galaxy clusters have been identified based on 
large photometric and/or spectroscopic surveys. In the Sloan Digital 
Sky Survey (SDSS) Stripe~82 region, a
zone covering a surface of 270~deg$^2$ across the celestial equator in
the Southern Galactic Cap ($-50 \degr < \alpha < 60 \degr$,
$|\delta|\leq 1.25 \degr$), several teams have found thousands of
candidate clusters by applying various methods. For example,
\citet{Geach11} detected 4,098 clusters up to redshift $z\sim 0.6$ 
by applying an algorithm that searches for statistically significant overdensities in a Voronoi tesselation of the projected sky.
\citet{Durret15} applied the AMACFI cluster finder (based on
photometric redshifts) in the same region and detected 3,663 candidate
clusters. In addition, galaxy cluster catalogues are constructed from 
the whole available SDSS survey area \citep[e.g.][]{Koester07, Wen09, 
Hao10, Szabo11, Rykoff14, Wen15}. 

The advantages of X-ray cluster surveys are first, that they provide almost
pure and complete cluster catalogues, and second, that the X-ray observables
(temperature and luminosity) correlate tightly with cluster masses
\citep[e.g.][]{Allen11,Boehringer13}. Several hundreds of clusters
were discovered in X-rays based on previous X-ray missions mainly from
ROSAT data \citep[e.g.][]{Ebeling98, Rosati98, Reiprich02,
  Boehringer04, Burenin07, Ebeling10}. 
  
The current X-ray missions
(Chandra, XMM-Newton, Swift/X-ray) provide contiguous cluster surveys
for small areas \citep[e.g.][]{Finoguenov07, Finoguenov10, Suhada12,
  Clerc14, Pierre15}, in addition to serendipitous cluster surveys in
the archival observations that provide a relatively large area of a
few hundred square degrees \citep[e.g.][]{Barkhouse06, Kolokotronis06,
  Lamer08, Fassbender11, Takey11, Mehrtens12, Clerc12, Tundo12,
  de-Hoon13, Takey13, Takey14}. So far, these surveys have provided a
substantial sample of several hundred clusters up to a redshift of
1.9 (photometric). An X-ray all-sky cluster survey will only be made
possible by the launch of eRosita, which will provide several hundred
thousand clusters \citep{Pillepich12}.

In the present paper, we present the 3XMM/SDSS Stripe 82 galaxy cluster
survey, a search for serendipitous galaxy clusters in the XMM-Newton
archive that are located in the footprint of Stripe 82 (S82
hereafter). The survey is based on extended detections in the third
XMM-Newton serendipitous source catalogue
\citep[3XMM-DR5,][]{Rosen15}. Since XMM-Newton has the largest
geometric effective area of any other X-ray satellite and a large
field of view of 30 arcmin, it provided a rich resource for
serendipitous cluster detection up to high redshifts. In addition, the SDSS
deep photometric data of S82 allowed estimating the photometric
redshifts of galaxies up to redshift unity \citep{Reis12}.

The main goal of our survey is to construct a cluster sample that can
be used for detailed multi-wavelength studies. The constructed sample
will include systems with a wide range of mass and redshift. This type
of group and cluster sample is needed to achieve our aim of
investigating the X-ray scaling relation (luminosity-temperature
relation) as well as optical properties (galaxy luminosity function
and morphological analysis of cluster galaxies).  Another product of
the present project is to develop a new method for identifying the optical
counterparts around the cluster centres determined from X-rays or from
the SZ-effect and for constraining their redshifts. This method will be based 
on a machine-learning technique \citep{Mahmoud16}.

We present here our strategy of selecting the X-ray cluster candidates in
our survey. As a first result, we provide a sample of optically
confirmed clusters, with their physical properties (redshift,
luminosity, and mass) and a list of the remaining X-ray cluster
candidates that can be used for further explorations or follow-up
programs.  
We also use the  spectroscopic redshifts of galaxies available from
the SDSS-DR12 to spectroscopically confirm the constructed cluster sample.

  As a second result, we report the discovery of two pairs of galaxy
  clusters that are strong candidates for merging clusters. Their
  properties derived from the available X-ray data are presented,
together with galaxy density maps. This is
  done to show the distribution of cluster galaxies and to determine
  the significance levels of structures detected in the cluster regions.
  The observation of binary clusters caught just before, during, or
  just after a collision are of particular interest. We can use such
  systems as probes of the cluster--cluster interaction, study how
  these powerful (albeit with Gyr duration) events affect the
  intra-cluster gas physics in mergers \citep[e.g.][]{Markevitch07,
    Bulbul16}, determine physical properties of the missing baryons in
  the form of filaments \citep[e.g.][]{Planck13a}, and analyse the
  interplay between galaxy members and clusters
  \citep[e.g.][]{Hallman04} and even the distribution of dark matter
  \citep{Clowe06}. We intend to obtain deeper Chandra and/or
  XMM-Newton data for these two pairs of galaxy clusters to perform
  this type of analysis. 

The structure of the paper is as follows. We first present the
selection procedure of the X-ray cluster candidates in Sect.~2. The
confirmed cluster sample and X-ray parameters are presented in
Sects.~3 and 4, while the remaining cluster candidates and their 
classifications are given in Sect. 5. The discovery of two new candidates 
for merging galaxy clusters is presented in Sect. 6. We summarise the 
work in Sect. 7.
Throughout this paper, we used the cosmological parameters
$\Omega_{\rm M}=0.3$, $\Omega_{\Lambda}=0.7$ and $H_0=70$\ km\
s$^{-1}$\ Mpc$^{-1}$.



\section{X-ray cluster candidates}

The 3XMM/SDSS Stripe 82 cluster survey is mainly based on the XMM X-ray serendipitous source catalogue. 
The latest version of the catalogue, 3XMM-DR5, was released on 2015
April 28. The 3XMM-DR5 catalogue contains 565,962 X-ray detections
comprising 396,910 X-ray sources, which were detected in 7,781 EPIC
(PN, MOS1, MOS2) observations made public on/or before 2013 December
31. These observations cover 877 deg$^2$ of the sky \citep{Rosen15}.

Our procedure to construct a cluster (candidate) sample has two
steps, as described below. We first select suitable XMM observations
for our survey, followed by selecting X-ray cluster candidates detected
in those fields.


\subsection{Selection of XMM-Newton observations}

The list of all XMM observations along with the parameters that were
used in the construction of the 3XMM-DR5 catalogue were made public by
the XMM Survey Science Centre (SSC) team at the catalogue
homepage\footnote{\url{http://xmmssc.irap.omp.eu/Catalogue/3XMM-DR5/3XMM_DR5.html}}.
To include an observation from this list in our survey, the
observation must fulfil the following criteria:

\begin{enumerate}

\item The observation must be completely covered in the S82 survey scans, meaning that the whole field of view is embedded in the area of S82 with boundaries of $-50.0 \degr  \le RA \le 60.0 \degr $ and $-1.25 \degr  \le Dec \le 1.25 \degr $ \citep{Annis14}.   

\item It is considered as a clean observation according to the
  observation class ($OBS\_CLASS$) parameter given in the 3XMM-DR5
  catalogue. This parameter indicates the fraction of the field of view
  that has been flagged (masked) in zones where the source detection and/or  
  characterisation are suspect \citep{Rosen15}. Here we selected
  detections with $OBS\_CLASS$ = 0 (nothing has been flagged), 1 ($ 0
  \% < flagged\ area < 0.1 \%$), and 2 ($0.1 \% \le flagged\ area < 1
  \%$).

\item The observation must be performed in the full frame mode with
  the full field of view exposed (prime full window, PFW) by at least
  EPIC PN or one of the MOS cameras.

\end{enumerate}

By construction, the selected XMM fields are located at high galactic
latitudes, $b < -35 \degr$, since the S82 region is chosen to be in the 
Southern Galactic Cap at  $b < -23 \degr$.  
This will minimise the effect of galactic absorption of the X-ray emission 
for the cluster sample. The
EPIC cameras were operated with thin and medium filters for 76 and 24
percent, respectively. This will maximise the scientific returns by
less affecting the effective area of the telescope, especially in the
soft X-ray band.

We limited our survey to those XMM fields that are located in S82
since its imaging is deeper by 2 magnitudes than other SDSS areas,
allowing the spectroscopic follow-up of galaxies in the SDSS-III
project to reach a redshift of 0.8. In addition, other data are 
available in the infrared (WISE, UKIDS) and at radio wavelengths
(FIRST). These available data will enable us to conduct multi-wavelength
studies of a large portion of the constructed sample.

Applying these selection criteria yielded 76 observations, which were
visually inspected through the web-page of the XMM observations and
data processing
status\footnote{\url{http://xmm2.esac.esa.int/external/xmm_mission_plan/index.php}}. This
led to the exclusion of only two observations that have high background, as
is clearly seen in the images and stated in the quality report given on
that web-page. As a result, 74 observations are considered in our
survey, their list is given in the Appendix A,
Table~\ref{tab:obsids-list}. The total area of these 74 fields is
11.25 deg$^2$, taking the overlap areas among the
fields into account.

\begin{figure}
  \resizebox{\hsize}{!}{\includegraphics{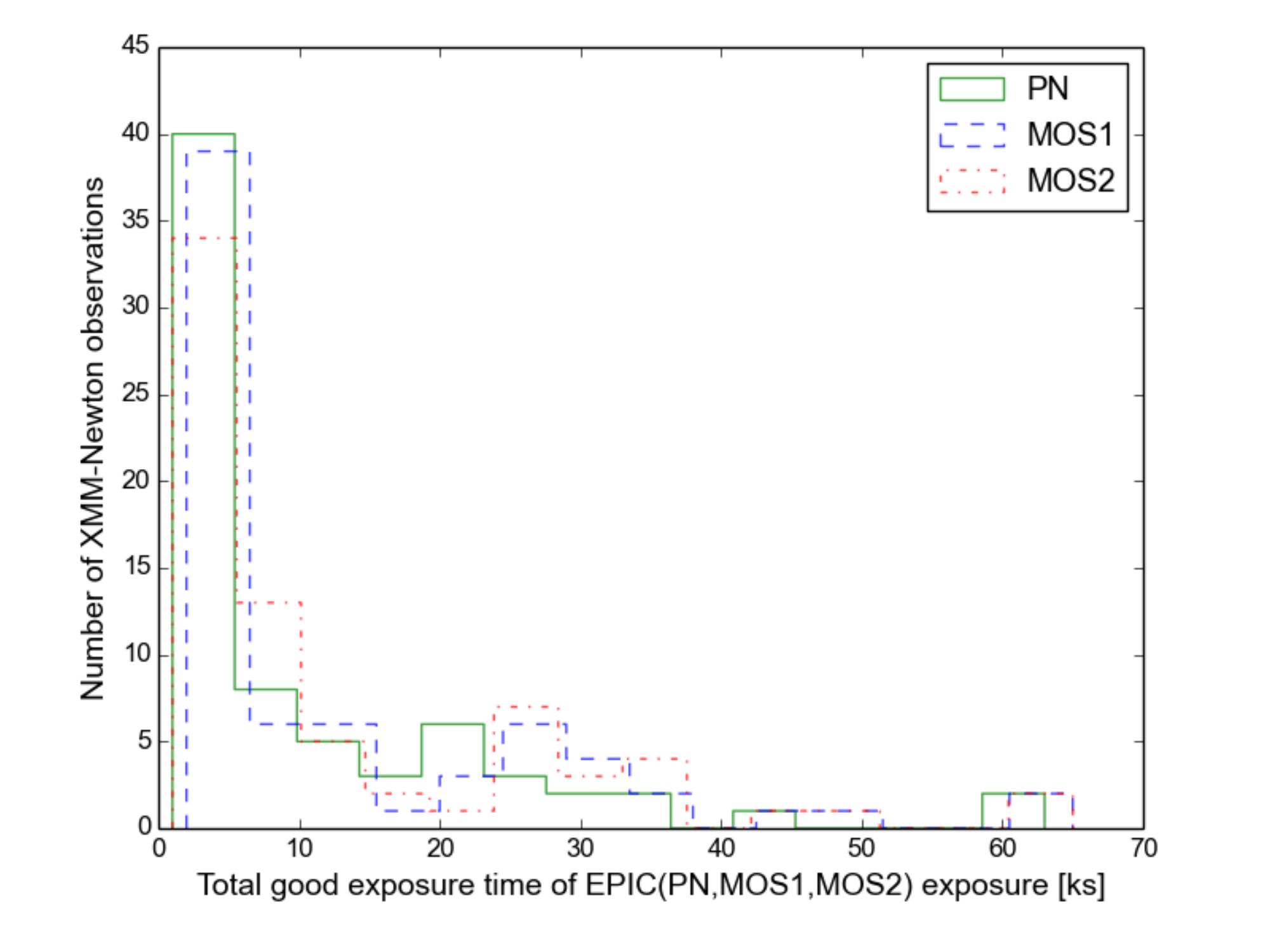}}
  \caption{Histogram of the EPIC (PN: green solid line, MOS1: blue dashed line, MOS2: red dash-dotted line) total good exposure time (after event filtering) of the survey fields (74 observations).} 
  \label{f:EPIC_TEXP}
\end{figure}

Since we did not apply a criterion on the EPIC exposure time, the list
of selected observations spans a wide range of good exposure times 
(good time intervals) from 2 ks to 65 ks. These exposures were given 
in the list of XMM observations that
was used to construct the 3XMM-DR5 catalogue. Figure~\ref{f:EPIC_TEXP}
shows the histogram of EPIC (PN, MOS1, MOS2) total good exposure times 
of the
fields in our survey. About half of the fields with exposure times
shorter than 5 ks were observed in a mosaic mode in the framework of the
XMM-Newton survey of Stripe 82. This leads to almost twice the
exposure time in the overlap regions as a result of the strategy of the mosaic
mode observations \citep{LaMassa13}.

\subsection{Selection of X-ray extended sources}

Galaxy clusters can be simply identified at X-ray wavelengths as X-ray
luminous, continuous, spatially extended, extragalactic sources
\citep{Allen11}. Therefore, the list of extended sources in the XMM
serendipitous source catalogue is considered as a valuable data base
to construct X-ray selected cluster samples. In the 3XMM-DR5
catalogue, the detection is considered as an extended source when the
extent radius is in the range of 6-80 arcsec. The extent radius is
determined by the Science Analysis System (SAS) task {\tt emldetect} 
that is derived by fitting a convolution of a $\beta$-model and a
point spread function (PSF) to the source distribution of counts
\citep{Rosen15}.

We selected all the extended detections from the 3XMM-DR5 catalogue
that are detected in the EPIC images of the 74 observations considered in our survey. This yielded a list of 120 extended
detections. This list includes repeated detections that are due to the overlap
between some observations. The 3XMM-DR5 catalogue provides a source ID
(SRCID) for each detection, which is a unique ID for multiple
detections. By using the SRCID as an indicator of a unique
source, the list of X-ray extended sources comprises 114 objects. 
We have the same number of unique sources if we consider two
detections as one source when their angular separation is smaller than
the sum of their core radii (EPIC extent parameters) and larger than
20 arcsec. If there is more than one detection for the same source, we
select the one with the highest photon counts. The multiple detections
available for the same source will be later used in the spectral
fitting to derive more accurate parameters.

Since we did not apply any criterion on the quality of the
detections, we visually inspected each source in two steps to exclude
spurious detections. The first step was to investigate
the X-ray images through the FLIX upper limit server\footnote{\url{http://www.ledas.ac.uk/flix/flix_dr5.html}}. This inspection enabled
us to remove sources that were located in the wings of bright sources, within a
large extended source, or at a large off-axis angle. Another visual
inspection was checking the corresponding colour images through
the navigation tool of the SDSS
server\footnote{\url{http://skyserver.sdss.org/dr12/en/tools/chart/navi.aspx}},
which allowed us to remove X-ray sources corresponding to the emission
of nearby bright galaxies or from star clusters.

As a result, after these visual inspection processes the list of X-ray
cluster candidates comprises 94 systems instead of 114 objects. Since
our survey covers an area of 11.25 deg$^2$, the spatial number density
of the candidate list is 94/11.25=8.36 candidates deg$^{-2}$. This is a
comparable spatial density to that determined
in the XMM-Newton Cluster Survey (XCS), 3675/410=8.96 candidates (with
$ > 50 $ background-subtracted counts) deg$^{-2}$
\citep{Lloyd-Davies11}.

\begin{figure}
  \resizebox{\hsize}{!}{\includegraphics{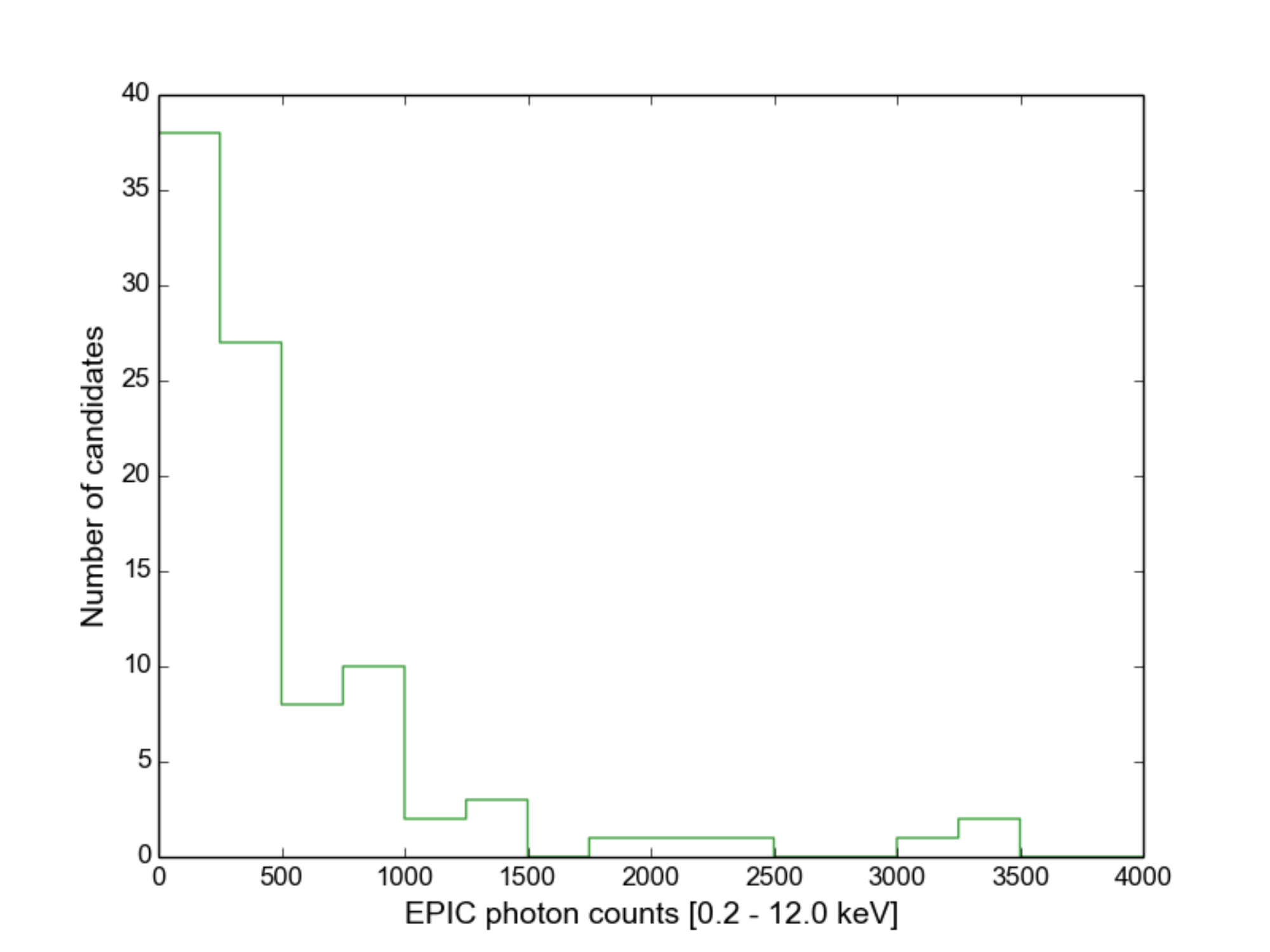}}
  \caption{Histogram of the EPIC photon counts in the broad energy
    band [0.2 - 12.0]~keV as given in the 3XMM-DR5 catalogue for the
    X-ray cluster candidate sample.}
  \label{f:hist_EP_8_CTS}
\end{figure}

The cluster candidate sample has a wide range of EPIC photon counts in
the energy band [0.2 - 12.0]~keV from 84 to 3440 counts, as shown in
Fig.~\ref{f:hist_EP_8_CTS}.  The reliability of detected sources in
the 3XMM-DR5 catalogue is evaluated by the maximum likelihood
parameter ($Det\_ML$), which varies from 6 (low reliability) to a few
million (high reliability). For our candidate sample, the EPIC
$Det\_ML$ parameter in [0.2 - 12.0]~keV has a range of 8 - 1949 with a
median of 35. We also note that the targets of the XMM fields are not
excluded in the current survey. Therefore, the EPIC off-axis angles of
the sample are smaller than 14 arcmin with a median of 7 arcmin.  
  We point out here that our X-ray cluster candidate list will not define
  a complete cluster sample since we did not apply an X-ray flux
  limit. In addition, X-ray extended sources in the 3XMM-DR5 catalogue
  are detected from observations with various exposure times, as shown
  in the previous subsection.

\section{Known clusters among the candidates}

To analyse the X-ray and optical data of the candidate sample, we determined the cluster redshifts. This can be done either by
using X-ray spectral fitting or optical imaging and/or
spectroscopy. X-ray redshifts can be estimated for candidates with
more than 1000 counts (12 percent, as shown in
Fig.~\ref{f:hist_EP_8_CTS}), 
\citep[see for example][]{Lamer08, Yu11, Lloyd-Davies11}. 
The alternative and main way is to estimate the cluster redshift based on optical/NIR photometric and spectroscopic data. 

To do this, we cross-correlated the X-ray cluster candidate
list with six available cluster catalogues selected in X-rays and/or
optical bands to identify the previously known clusters and to
obtain their redshifts, as described below. To complement the list of
known clusters, we searched the NASA Extragalactic Database (NED) to
obtain redshifts for systems that were not in the cross-matched
catalogues. We then present the whole list of previously known clusters with redshift measurements as well as the available spectroscopic redshifts for cluster galaxies from the SDSS-DR12. 


\subsection{X-ray cluster surveys}

\cite{Takey11} presented a similar survey based on X-ray extended
detections selected from an earlier version of the XMM-Newton
serendipitous source
catalogue \cite[2XMMi-DR3\footnote{\url{http://xmmssc-www.star.le.ac.uk/Catalogue/xcat_public_2XMMi-DR3.html}},][]{Watson09}
located in the SDSS-DR7 sky region. The 2XMMi-DR3 catalogue
was based on XMM observations made before October 2009.  They
presented a cluster sample of 530 systems based on photometric
redshifts of galaxies in SDSS-DR8 and another sample of 383 galaxy
groups and clusters based on spectroscopic redshifts of galaxies in
SDSS-DR10. In total, they have estimated the redshift of 597 X-ray
selected clusters in the range of 0.03-0.79. Of these, 455 clusters
have spectroscopic confirmation for at least one member galaxy with a
spectrum in SDSS \citep{Takey13, Takey14}.

\cite{Mehrtens12} presented the first data release of the XMM-Newton
Cluster Survey (XCS), which was based on the entire XMM-Newton archive
available till July 2010. The catalogue consists of 503 optically
confirmed galaxy groups and clusters, among which 463 systems with redshift
measurements in the range of 0.06-1.46. About half of these sources
have spectroscopic confirmation either from the literature, follow-up
campaigns, or SDSS.

Another serendipitous cluster catalogue was constructed in the framework 
of the XMM Cluster Archive Super Survey (X-CLASS) by
\cite{Clerc12}. The catalogue comprises 422 cluster candidates
with high signal-to-noise ratio (S/N)
purely based on X-ray criteria from available XMM
archival observations in or before May 2010. About half of these
candidates are optically confirmed and assigned redshift estimates in
the range from 0.02 to 1.73. The number of galaxy clusters with
spectroscopic confirmation is 78, as given in their public catalogue.

We cross-matched our cluster candidate list and the X-ray
selected cluster samples with redshift estimates from these three
cluster surveys within a matching radius of 20 arcsec. In total, about
one third of our candidates were selected and published in these
surveys. The statistics of these X-ray cluster surveys and the matched
cluster samples are given in Table~\ref{tab:cross-matchin}.

  We note that some clusters in the matched catalogues
  are located in the S82 region, but are not included in the common
  clusters resulting from the cross-matching process. There are
30, 40, and 6   X-ray galaxy clusters in S82 from the 2XMMi/SDSS, XCS, and X-CLASS
  catalogues, respectively. The clusters
  might be missing because the respective observations are not considered
  in the 3XMM-DR5 catalogue and/or in our current survey, because
of different
  source detection algorithms, and/or because of our chosen cross-matching
  radius. Since the list of XMM observations used in these surveys is
  not public, the detailed comparison of recovering or missing
  clusters is not straightforward. A comparison between 422 extended
  sources detected in the X-CLASS catalogue and sources in the
  3XMM-DR5 catalogue was reported by \citet{Rosen15}. They found that
  59 of 422 sources (in general, faint or irregular detections) were
  classified as point sources, while the remaining objects were
  detected as extended sources. This indicates that we might miss some
  clusters when their detections are classified as point sources in
  the 3XMM-DR5 catalogue.  
   

\subsection{Optical cluster surveys}

The largest published catalogue of optically selected galaxy clusters
so far, based on the SDSS data, was compiled by \cite{Wen12}. They
identified 132,684 clusters using the photometric redshifts of galaxies
in SDSS-DR8. The catalogue was updated to re-estimate the cluster
richness and to include the new spectroscopic redshifts available in
SDSS-DR12. They also constructed a cluster catalogue comprising 25,419
new rich high-redshift clusters by modifying the cluster detection
algorithm and using SDSS-DR12 \citep{Wen15}. We combined the
two catalogues in one table comprising 158,103 systems in the redshift
range of 0.05-0.8 (WH15 hereafter), which was used in the
cross-matching process. About 77 percent of these systems have
spectroscopic redshifts for at least one galaxy member from SDSS as
listed in both catalogues.

Another cluster survey based on SDSS-DR8 data is the redMaPPer
red-sequence cluster finder by \cite{Rykoff14}. The catalogue
comprises 26,350 clusters across the redshift range of 0.08-0.55. Of
these clusters, 16,259 BCGs (62 percent) have spectroscopic redshifts
from SDSS data, as given in their catalogue.

\cite{Geach11} presented a cluster catalogue of 4,098 objects
identified in the SDSS Stripe 82 region based on statistical galaxy
overdensities. The redshift of these systems ranges from 0.04 to
0.8. About 32 percent of these clusters have spectroscopic redshifts
for at least one member galaxy from public spectroscopic data in
SDSS-DR7, 2SLAQ and WiggleZ. Unfortunately, these spectroscopic
redshifts are not given in their catalogue.

The positions of the X-ray cluster candidates are matched to these
three catalogues using a search radius of one arcmin. The optically selected clusters are centred on the positions of the brightest cluster galaxies (BCGs), which are located mostly within 150 kpc from the X-ray positions \citep{Takey13}. Therefore, the matching radius of one arcmin is suitable for the expected cluster redshift range of our sample $z > 0.05$. This matching process provides an
identification of optical counterparts and redshift estimates for 41
percent of the cluster candidates. When a spectroscopic redshift in these catalogues was available for the optical counterpart,
we used it instead of the photometric
one. Table~\ref{tab:cross-matchin} lists the statistics of the matched
cluster catalogues and the resulting common systems from the
cross-matching process.


\subsection{Literature redshifts: NED}

About half of the X-ray cluster candidates were previously identified
from both X-ray and optical cluster surveys, as mentioned above. To
obtain redshift values for more cluster candidates, we searched the NED
for available redshifts. If an optical or X-ray detected cluster was
found in NED with an offset of less than one arcmin from the X-ray
emission peak, we assigned its redshift to the cluster candidate. 
In this way, we ensured that the redshift values obtained from the
cross-matching process are correct since some of them were confirmed
from other surveys or by checking cluster galaxies with spectroscopic
redshifts listed in NED. In addition, four cluster candidates with
redshift measurements were detected in different projects
\citep{Wen09, Gerke12, Menanteau13}, as stated in
Table~\ref{tab:cross-matchin}.
 

\subsection{Cluster sample with redshift measurements}

\begin{table}
 \caption{\label{tab:cross-matchin} Known clusters of galaxies (CLG) with redshift measurements for the candidates in the literature.}
 \centering
 \begin{tabular}{l c c c c }
  \hline\hline
  CLG        & Nr. CLG   & Redshift  & Matching &  Nr. CLG\\
  catalogue  & catalogue & range     & radius ($'$) &  common \\
  \hline
  2XMMi/SDSS & 597       & 0.03-0.79 & 0.3  & 25 \\
  XCS        & 503       & 0.06-1.46 & 0.3  & 19 \\
  X-CLASS    & 422       & 0.02-1.73 & 0.3  & 5 \\
  \hline 
  WH15       & 158,103   & 0.05-0.8  & 1.0  & 31 \\
  redMaPPer  & 26,350    & 0.08-0.55 & 1.0  & 12 \\
  Geach      & 4,098     & 0.04-0.8  & 1.0  & 28 \\
  \hline
  \multicolumn{3}{l}{NED (CLGs are not in the matched catalogs)} & 1.0 &  4 \\
  \hline
  \multicolumn{3}{l}{Total (unique cluster sample)}&   &  54 \\
  \hline
 \end{tabular}
\end{table}

The cross-matching of the X-ray cluster candidates with the six 
mentioned optical and X-ray cluster catalogues and searching NED
yielded 54 clusters with redshift measurements, as listed in the last
row of Table~\ref{tab:cross-matchin}. Of these 54 systems, 35 clusters
are entered in more than one of the catalogues matched to ours. In
these cases, we took the redshift from the catalogue that provides a
spectroscopic redshift. About 83 percent of the cluster sample are
spectroscopically confirmed according to the information given in the
mentioned catalogues or listed in the NED.

The X-ray emission of one third of the cluster sample with redshift
estimates is detected for the first time, while the remainder are known
in one or more X-ray cluster surveys.
It is worth pointing out that the current cluster survey includes some
recent XMM observations that had not been performed or made public
when conducting the previous XMM cluster surveys (XCS, 2XMMi/SDSS,
X-CLASS). These surveys included XMM observations until July 
2010 at the latest. Therefore, we explored more XMM-Newton fields (40 observations)  located in the Stripe 82 made from July 2010 until December 2013. 
The known X-ray clusters from the cross-matching process are marked 
in the column ``X-ray Catalogue'' of Table~\ref{tab:CLG-cat} by their 
catalogue name, while the new X-ray clusters have no entries in the 
column and the space is filled by $``$-9999$''$.

\begin{figure}
  \resizebox{\hsize}{!}{\includegraphics{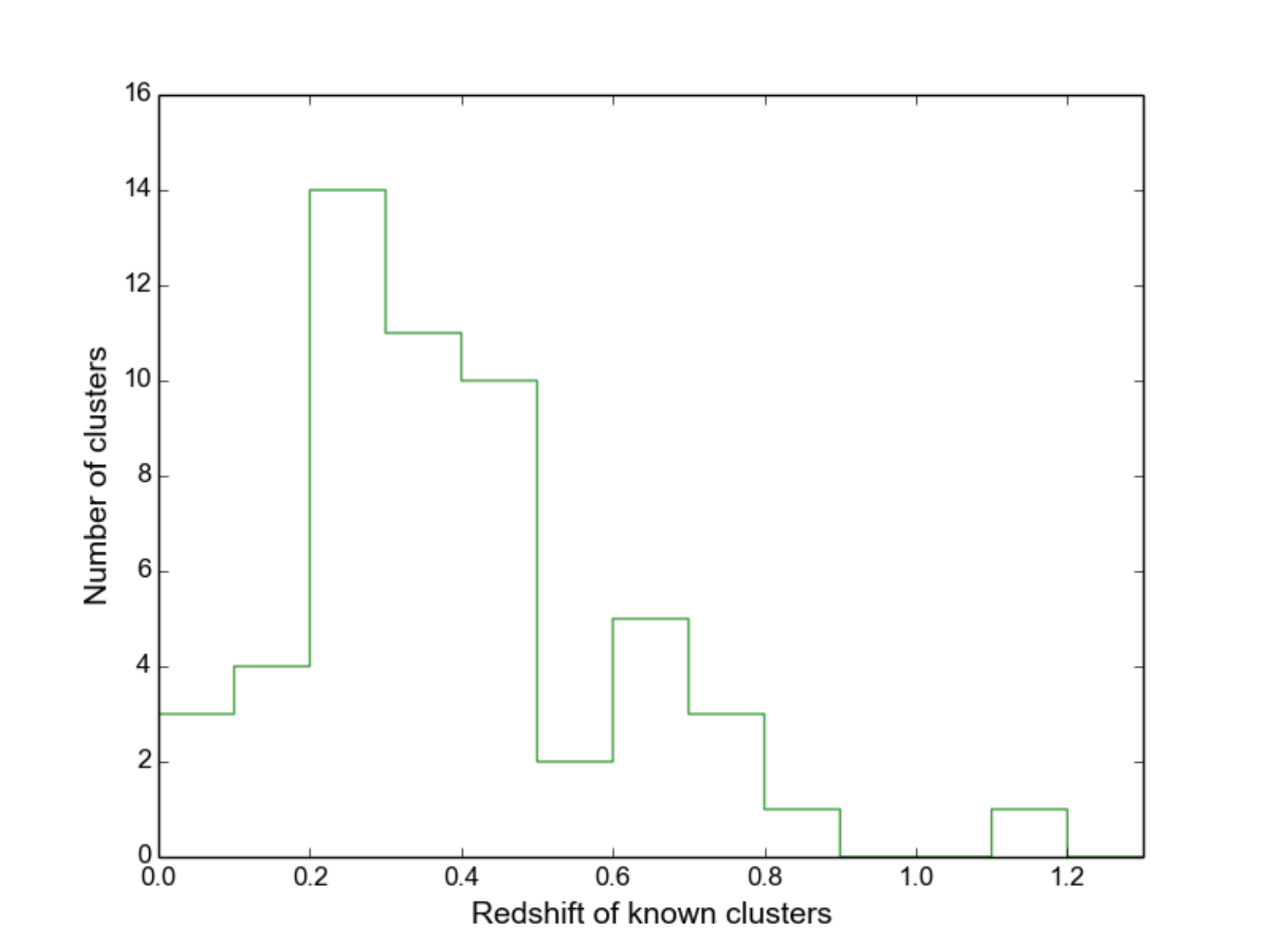}}
  \caption{Histogram of the redshift distribution of our sample of 54
    clusters that were previously known in the literature.}
  \label{f:hist_z_CLG}
\end{figure}

Our cluster sample spans a broad redshift range from 0.05 to 1.19. We
note here the re-detection of a cluster at redshift of 1.19, which was
first discovered by \cite{Dietrich07}, spectroscopically confirmed
by \cite{Suhada11}, and listed in the XCS cluster catalogue. This is
the only source in common with the distant cluster sample from the
XMM-Newton Distant Cluster Project (XDCP) published by
\cite{Fassbender11}. Our cluster sample also includes ten systems at
high redshift beyond 0.6. Figure~\ref{f:hist_z_CLG} shows the redshift
distribution for the previously known cluster sample.

\begin{figure}
\centering
   \resizebox{7 cm}{!}{\includegraphics[viewport= 11 8 384 381, clip]{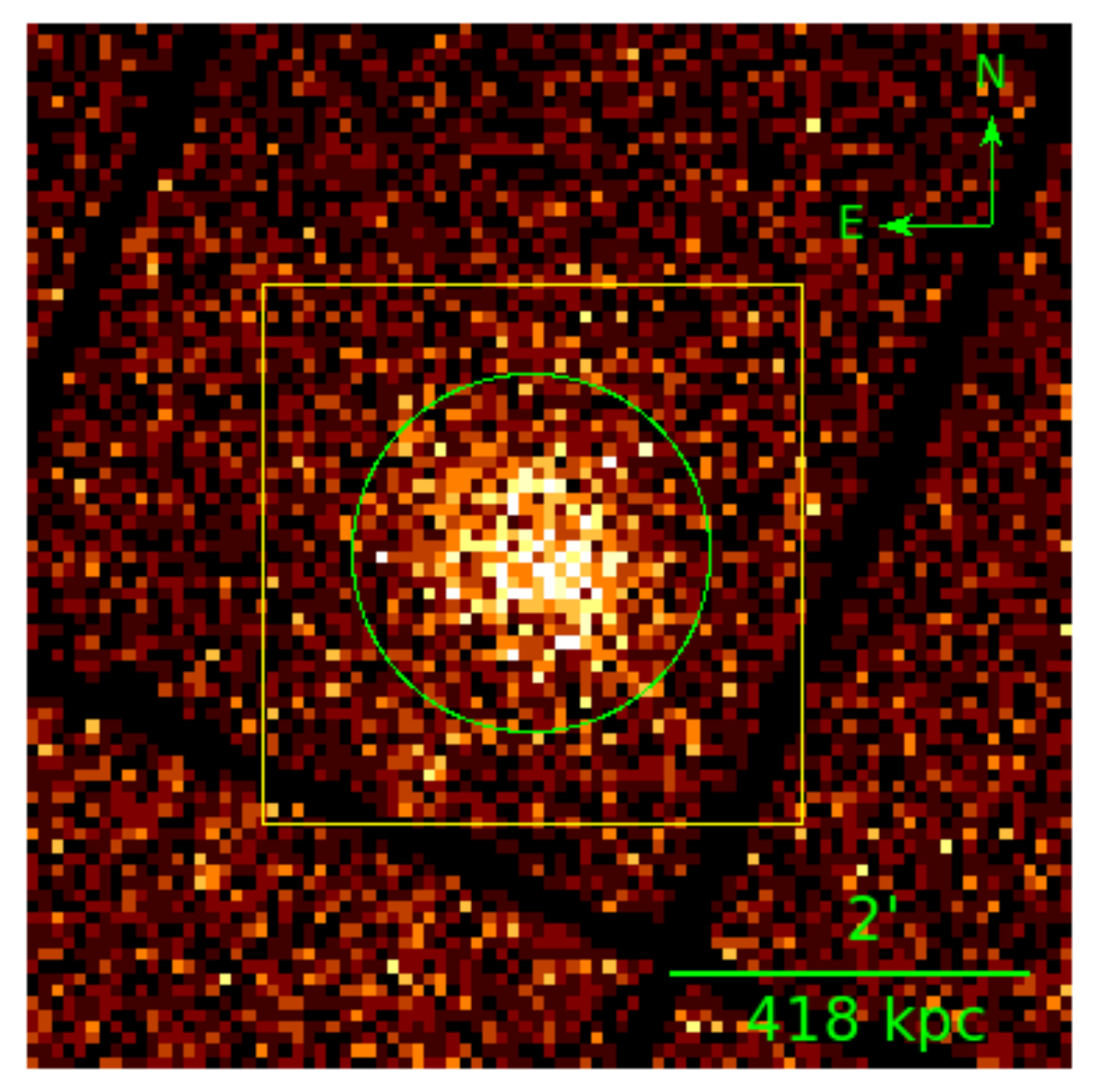}}
  \resizebox{7 cm}{!}{\includegraphics{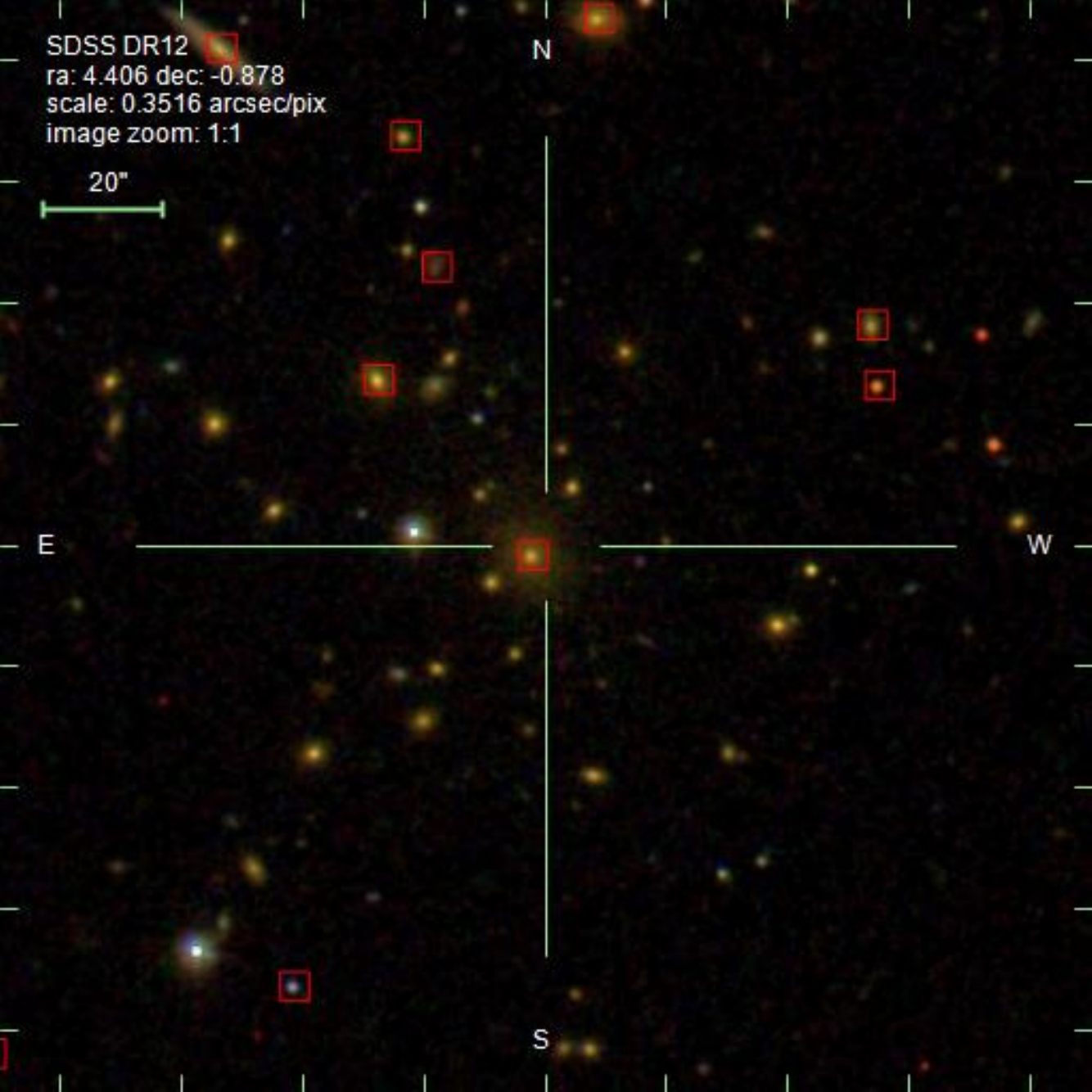}}
  \caption{Example cluster (DETID: 104037601010003, IAU Name: 3XMM
    J001737.3-005240) at a redshift of 0.2141 (spectroscopic
    redshift). Upper panel: X-ray image (EPIC PN from the pipeline products, $6'\times 6'$) in
    [0.2-12.0]~keV of XMM-Newton observation (0403760101) targeted at a
    starburst galaxy. The green circle (radius=$1'$) and yellow box
    ($3'\times 3'$) are centred on the X-ray emission peak. Lower
panel: SDSS colour image corresponding to the yellow box centred on the
    X-ray position as indicated by the cross-hair. Sources with
    spectroscopic redshifts from SDSS-DR12 are marked by red
    squares. }
  \label{f:104037601010003}
\end{figure}

The cluster sample list with X-ray positions and redshift estimates is
given in Table~\ref{tab:CLG-cat}, which is also available online at
the CDS.  Figure~\ref{f:104037601010003} shows the images of an example
cluster in X-rays and optical wavelengths. This is one of the X-ray
brightest clusters in our sample that was also detected in the XCS and
2XMMi/SDSS projects and identified by the three optical cluster
finders (WH15, redMaPPer, and Geach).

\subsection{Spectroscopic redshifts for our cluster sample from SDSS-DR12}

The cluster sample (54 galaxy groups/clusters) includes 45 and 9
systems that have spectroscopic and photometric redshifts from the cross-matched catalogues and/or from NED, respectively. We
explored the SDSS-DR12 to have spectroscopic confirmation for our
cluster sample, especially those systems with only photometric
redshifts. First, we extracted the parameters (ObjID, position,
magnitudes) of all galaxies surrounding the cluster centres (X-ray
positions) within 10 arcmin from the {\tt Stripe 82} table in the
Catalog Archive Server ({\tt CasJobs}) database \citep{Annis14}. 
Second, the galaxy sample was cross-matched with galaxies with 
spectroscopic redshifts in the {\tt SpecObj} table of SDSS-DR12 
\citep{Alam15}. Third, the photometric redshift of each galaxy 
was taken from the galaxy photometric redshift catalogue 
by \citet{Reis12}. 

To identify cluster galaxies with spectroscopic redshifts, we first
need to define a physical search radius and a redshift range for the
cluster redshift. We chose a search radius of $R_{200}$ since it is
comparable to the virial radius. In the next section, we estimated 
$R_{500}$ for each system. We converted the estimated $R_{500}$ to
$R_{200}$ by applying a factor of 1.5, which is derived as the mean
and median value of $R_{200}/R_{500}$ for 401 clusters in the XCS
cluster sample \citep{Mehrtens12}. $R_{500}$ and $R_{200}$ are defined
as the cluster radii within which the density is 500 and 200 times the
critical density of the Universe at the cluster redshift,
respectively. The range of cluster galaxy redshifts is $\pm
0.02(1+z_{\rm CLG})$ and $\pm 0.04(1+z_{\rm CLG})$ for clusters with
spectroscopic and photometric redshifts in our sample, respectively.

\begin{figure}
  \resizebox{\hsize}{!}{\includegraphics{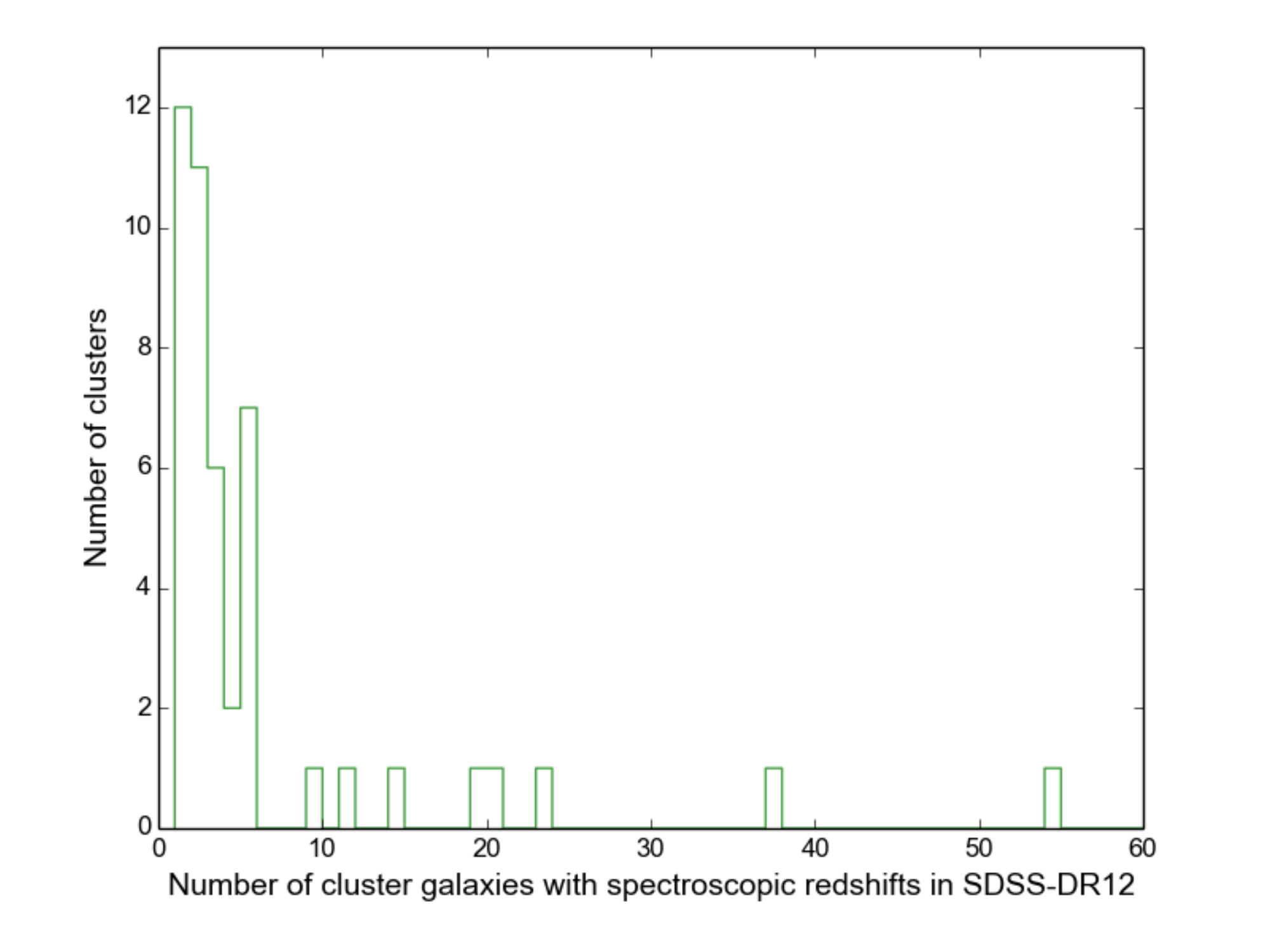}}
  \caption{Histogram of cluster galaxy members with spectroscopic
    redshifts available in the SDSS-DR12 that are consistent with the
    cluster redshift (see text). The bin size is one.}
  \label{f:hist_Nzs_DR12}
\end{figure}

As a result, 46 out of 54 clusters have spectroscopic confirmations based on
at least one member galaxy with a spectrum in SDSS-DR12. The remaining
systems are at high redshifts ($z \ge 0.65$) beyond the magnitude
limit of SDSS (6 clusters) or at low redshift ($z \le 0.46$) with no
available spectra (2 objects). Figure~\ref{f:hist_Nzs_DR12} shows the
distribution of cluster galaxies with spectroscopic redshifts within
$R_{200}$ from the X-ray emission peaks. In this way, we
spectroscopically confirmed six clusters that have only photometric
redshifts in the literature. There are still three clusters in our
sample that need to be spectroscopically confirmed. Table~\ref{tab:CLG-cat} lists in the last two columns the number of cluster galaxies with spectroscopic redshifts, when available, and the mean value of these redshifts.

\begin{figure}
  \resizebox{\hsize}{!}{\includegraphics{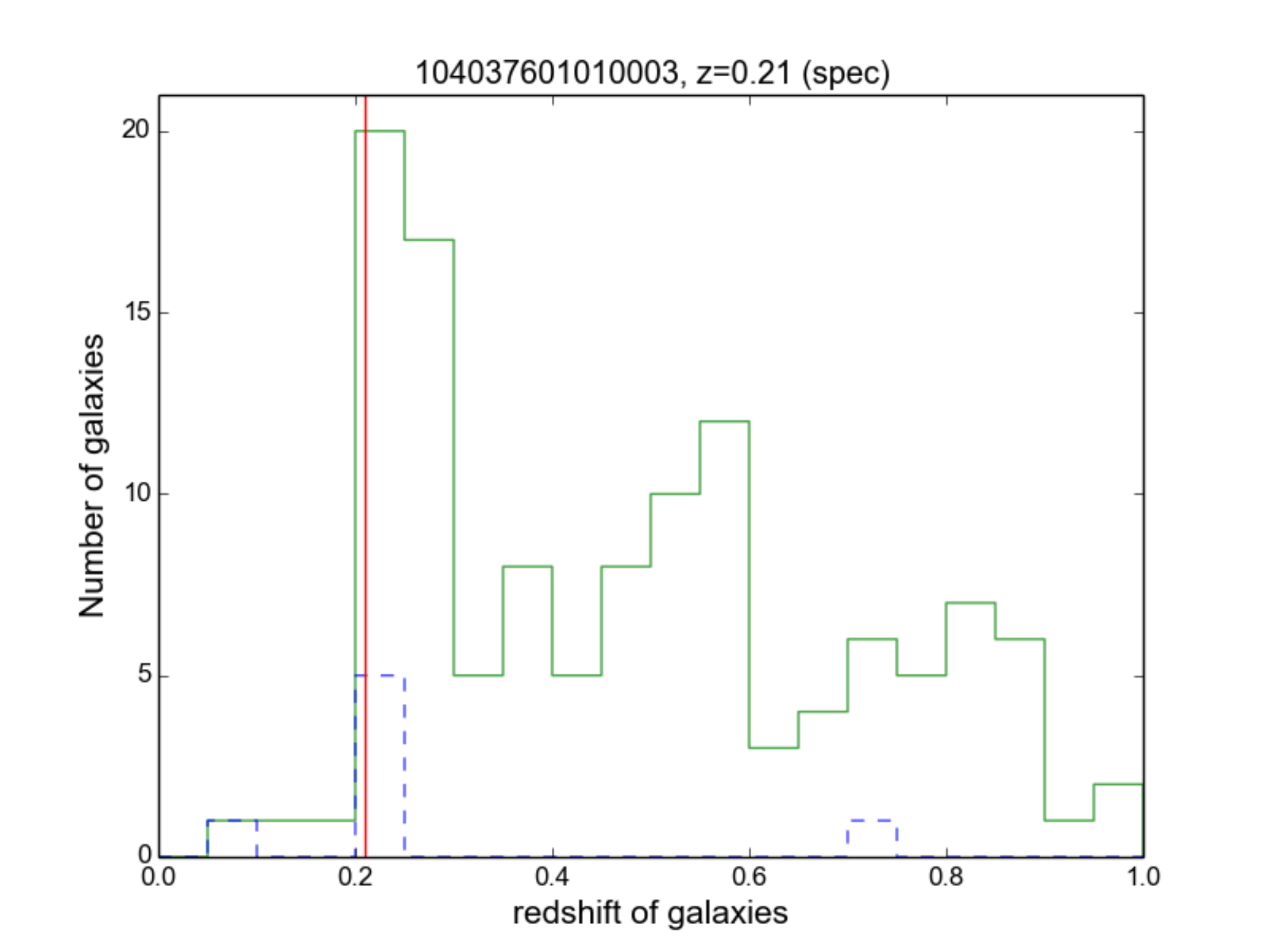}}
  \caption{Histogram of photometric (green solid line) and
    spectroscopic (blue dashed line) redshifts of galaxies located
    within 2 arcmin from the X-ray position of the example cluster
    (3XMM J001737.3-005240). The vertical red line indicates the
    spectroscopic redshift (0.2141) of the cluster that is consistent
    with the peaks in the redshift histograms.}
  \label{f:104037601010003_hist}
\end{figure}

  The previous step was to confirm the spectroscopic redshifts
  given in the cross-matched catalogues and to spectroscopically
  confirm some clusters that previously had only photometric
  redshifts. We also performed tests to validate the optical detections
  and the redshifts of our cluster sample. First, when a cluster was
  identified in two or more catalogues with comparable redshifts, we
  confirmed its detection and redshift. Fourteen, ten, seven,
and four clusters are entered
  in two, three, four, or five catalogues (the matched clusters), respectively.  In total, 35 systems are detected
  in at least two cluster surveys. For all clusters, we also searched
  NED for redshifts available from other projects, see Sect.
  3.3. Second, we plotted for each cluster the histogram of the
  photometric and spectroscopic redshifts of all the galaxies within 2
  arcmin from the X-ray emission peak, as shown in
  Fig.~\ref{f:104037601010003_hist} for the example cluster
  3XMM~J001737.3-005240. When a peak was consistent with the
  cluster redshift, we confirmed the cluster redshift, as for the
  example cluster. Third, we checked the spectroscopic redshift of the
  brightest galaxies near the X-ray positions through the SDSS
  skyserver tool, and also the spectroscopic redshifts available
  within $R_{200}$ from the X-ray centres.

  As a result of these tests, we confirm the redshifts given in our
  sample. Only in one case two clusters overlap at
  redshifts of 0.46 \citep[photometric, ][]{Szabo11, Takey13} and
  0.067 \citep[spectroscopic, ][]{Merchan05, McConnachie09}. We
  selected the cluster with the smallest physical offset between the BCG
  candidate and the X-ray emission peak. Two systems remain without
  spectroscopic confirmation.

\section{X-ray parameters of the known cluster sample with redshift measurements}

Reliable X-ray temperatures can be obtained from spectral fitting when
clusters have more than 300 net X-ray photon counts, as demonstrated by
\cite{Lloyd-Davies11}. The uncertainties of the other estimated
parameters (flux and luminosity) are also dependent on the brightness
of the cluster. Two thirds of our cluster sample with redshifts have
enough counts to derive X-ray temperatures with reasonable
uncertainties. We postpone the X-ray spectral analysis for the cluster
sample to an upcoming paper until we estimate redshifts for most of the
remaining X-ray cluster candidates.

In the present paper, we estimate the X-ray properties of the cluster
sample based on  data available in the 3XMM-DR5 catalogue and scaling
relations in the literature. The catalogue provides aperture-corrected
$\beta$-model fluxes in different energy bands for individual cameras
that were calculated with the SAS task {\tt emldetect}. The EPIC flux in
each band was computed as the weighted mean of the band-specific
detections in all cameras. The weight was based on the flux error
\citep{Rosen15}.
We determined the combined EPIC (PN, MOS1, MOS2) fluxes in the energy
band [0.5-2.0]~keV, $F_{\rm cat,\,0.5-2}$, as the sum of the EPIC
fluxes in the [0.5-1.0]~keV and [1.0-2.0]~keV bands. The error of
$F_{\rm cat,\,0.5-2}$ was determined as the propagated error from the
flux errors in the two bands. For the cluster sample, we computed the
X-ray luminosity and its error in the [0.5-2.0]~keV band, $L_{\rm
  cat,\,0.5-2}$, based on $F_{\rm cat,\,0.5-2}$, its error, and the
determined redshift.

\cite{Takey14} presented a relation (Fig. 11 and Eq.~1) that can
be used to convert the X-ray luminosity in the [0.5-2.0]~keV band,
based on the flux given in the 2XMMi-DR3 catalogue, into X-ray
bolometric luminosity ($L_{500}$) within a physical radius of
$R_{500}$. Since the fluxes in the 2XMMi-DR3 and 3XMM-DR5 catalogues
are not identical for a given source, we updated this relation
based on the new flux estimates given in the 3XMM-DR5 catalogue. As
stated in Sect. 3, 25 clusters are in common between the
cluster sample selected from the 2XMMi-DR3 and 3XMM-DR5
catalogues. These 25 common systems have estimates for $L_{500}$ based
on either X-ray spectroscopic parameters from spectral fitting
(15/25) or catalogue fluxes (10/25). We note here that we revised the redshift estimates for two clusters in the common sample.  

\begin{figure}
  \resizebox{\hsize}{!}{\includegraphics{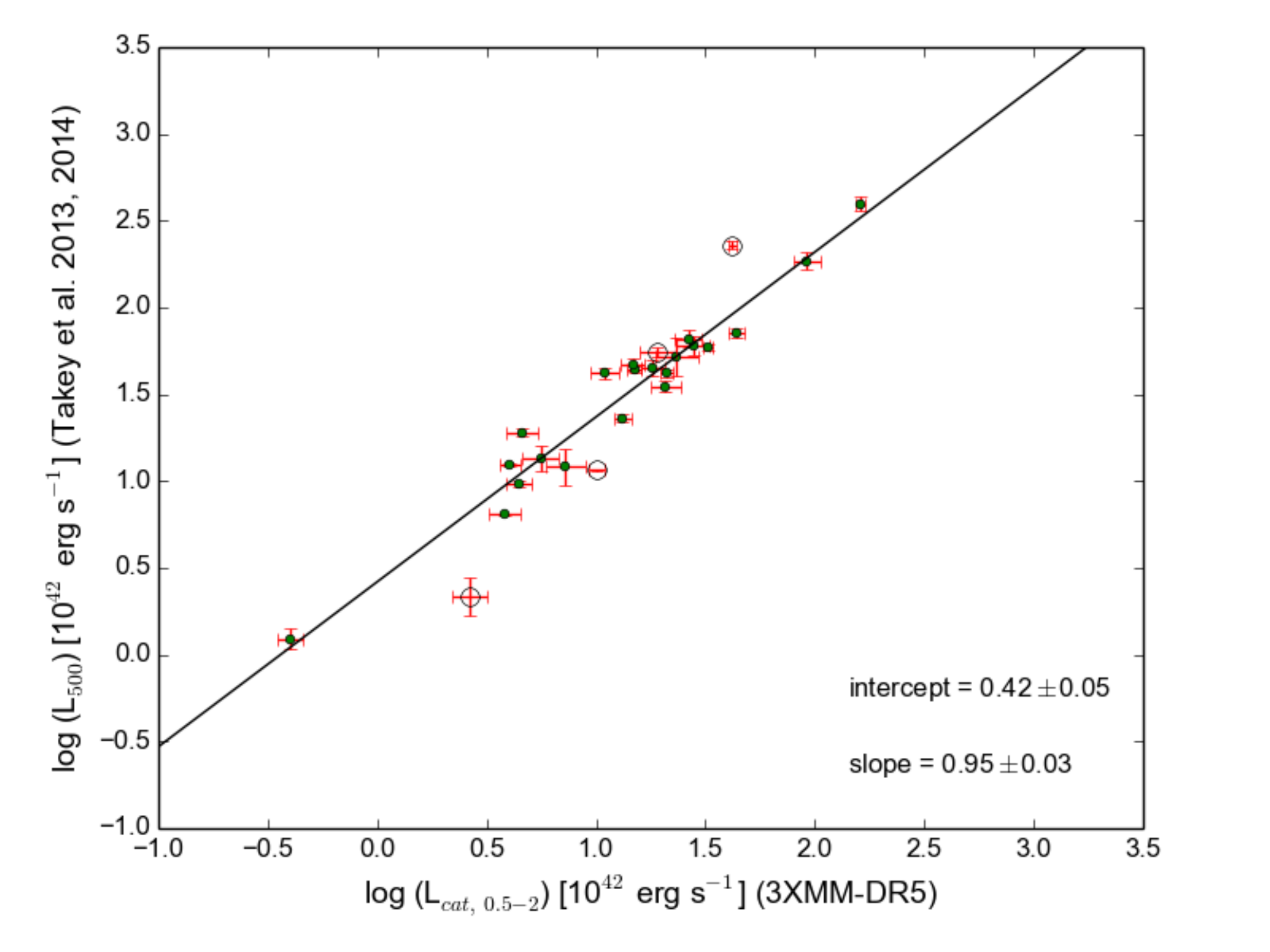}}
  \caption{Correlation between the X-ray luminosity $L_{\rm cat,\,0.5-2}$ based on the flux given in the 3XMM-DR5 catalogue and the X-ray bolometric luminosity $L_{500}$ derived by \cite{Takey13, Takey14} for the 25 common clusters. The solid line represents the best fit of the relation (BCES orthogonal regression, Python module) based on 21 systems (green dots) after excluding four systems (blue open circles), see the text for more detail. The best-fit parameters are written in the lower right corner.} 
  \label{f:L500_Lcat}
\end{figure}

For the systems in common except for the two clusters with revised
redshifts, we correlated $L_{\rm cat,\,0.5-2}$ from the current work
with $L_{500}$ from \citep{Takey13, Takey14}, as shown in
Fig.~\ref{f:L500_Lcat}. The slope and intercept of the relation,
derived using the BCES orthogonal regression method \citep[bces Python module,][]{Akritas96},
are $1.06 \pm 0.09$, $0.30 \pm 0.11$, respectively. The fitted parameters have large errors. This is due to the small number
of common systems and to the existence of two systems with a large
dispersion from the best-fit line. When we also exclude these two
outliers, the slope and intercept have smaller errors: $0.95 \pm 0.03$
and $0.42 \pm 0.05$, respectively. The best fit of the relation after
excluding four systems from the common systems is shown in
Fig.~\ref{f:L500_Lcat}, which is considered in the following
steps. The slope of the current relation $0.95 \pm 0.03$ agrees 
with the value of $0.91 \pm 0.02$ derived by \cite{Takey14}.

We used the relation shown in Fig.~\ref{f:L500_Lcat} to derive
$L_{500}$ for all the cluster sample from the estimated $L_{\rm
  cat,\,0.5-2}$. The errors of $L_{\rm cat,\,0.5-2}$, slope, and
intercept as well as the intrinsic scatter of the conversion relation
were considered when we estimated the error in $L_{500}$.  The
scaling relation presented here provides a simple way to convert $L_{\rm
  cat,\,0.5-2}$ into $L_{500}$ taking the bolometric
correction and the extrapolation of aperture corrected flux to $R_{500}$ into account.

\begin{figure}
  \resizebox{\hsize}{!}{\includegraphics{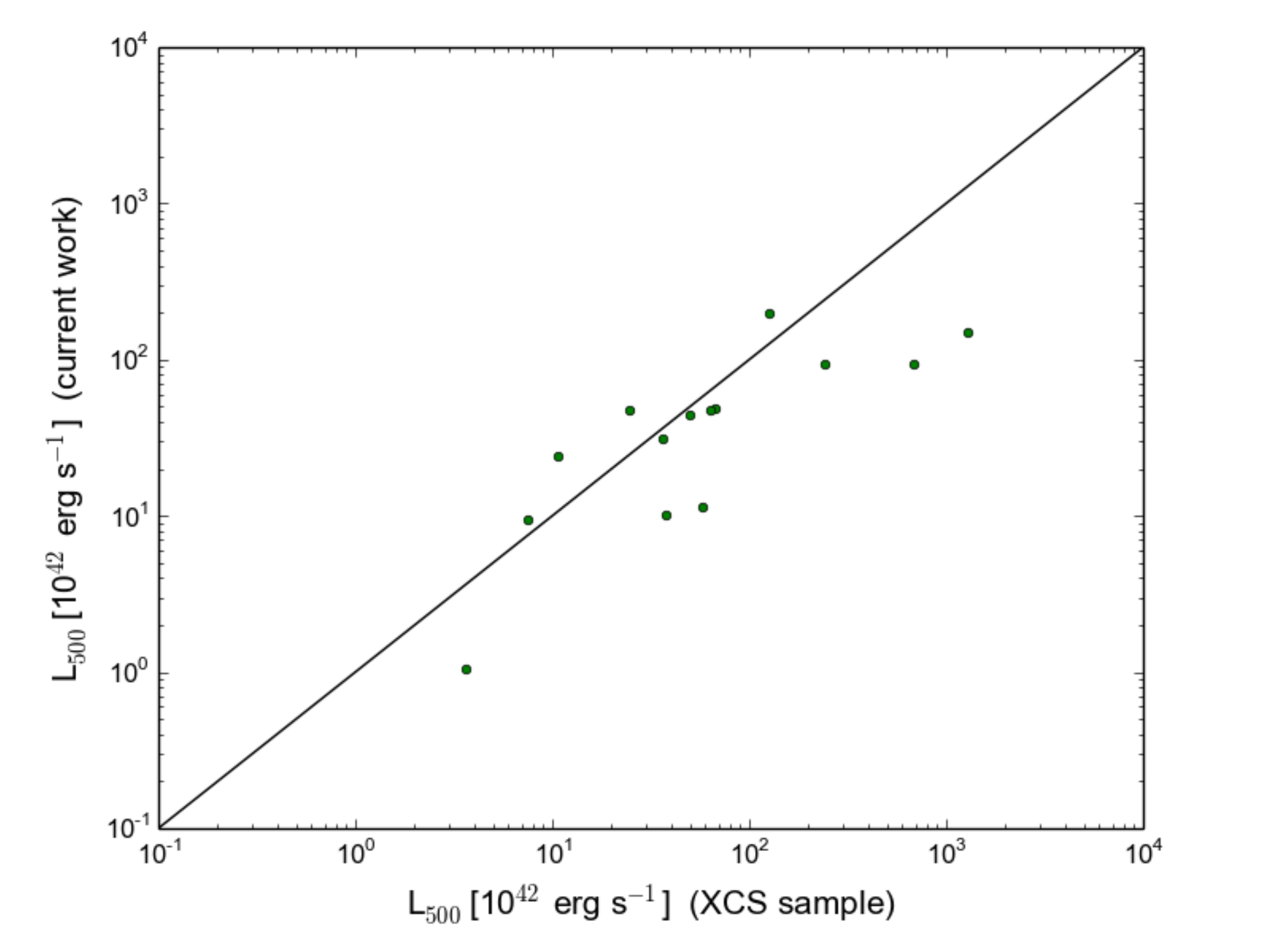}}
  \caption{Comparison of the current values of X-ray bolometric
    luminosity $L_{500}$ and those derived in the XCS project for
    14 common systems. The solid line represents the diagonal of the
    square.}
  \label{f:L500_Lxcs}
\end{figure}

To ensure that the conversion relation is acceptable, we compare the
estimated luminosities with the values available in the literature. As
mentioned in Sect. 3, there are 19 common systems between the
current cluster sample and the XCS sample. Of these, only 14 clusters
have $L_{500}$ estimates in the XCS project. Figure~\ref{f:L500_Lxcs}
shows the comparison between the $L_{500}$ measurements from XCS and
current work, which are in  fair agreement for this  small
sample. We should mention that three of the common systems used in the
comparison have revised redshifts (from photometric to
spectroscopic ones) in the current project. In addition, different procedures 
were used to compute $L_{500}$ in the two projects. Moreover, the estimated $L_{500}$ is mainly based on the flux given in the 3XMM-DR5 catalogue. If the flux estimate is inaccurate, this will derive an inaccurate estimate of $L_{500}$ as well. 
The cluster sample has 
X-ray bolometric luminosities $L_{500}$ in the range $(1 \rm - \rm 356)
\times 10^{42} \rm erg\ \rm s^{-1}$ with a median of $35.2 \times
10^{42} \rm erg\ \rm s^{-1}$.

The estimated $L_{500}$ was used to determine the cluster mass,
$M_{500}$, using the $L - M$ relation derived from a representative
local cluster sample (REXCESS) by \citet{Pratt09}, assuming the 
relation is valid for distant clusters as well. We then determined
$R_{500}$ based on the derived value of $M_{500}$. The majority of the
cluster sample is in the low- and intermediate-mass regime since the
range of masses is $(2 \rm - \rm 25) \times 10^{13}$\ M$_\odot$ and
the median is $8.4 \times 10^{13}$\ M$_\odot$. In the cluster
sample are only three systems in the ROSAT cluster sample
\citep{Piffaretti11}. The mass measurements of these three systems agree well with the values derived from ROSAT data.

Table~\ref{tab:CLG-cat} lists the cluster sample that comprises 54
systems with their redshift measurements as well as their X-ray
properties. The catalogue is also available in full form at the
CDS. The columns of the catalogue are listed in Table B.1 of Appendix B.

\section{Sample of cluster candidates}

We classified the remaining X-ray cluster candidates (40 systems) without
available redshift estimates in the literature based on the appearance in X-ray images. We used the FLIX upper
limit server\footnote{\url{http://www.ledas.ac.uk/flix/flix_dr5.html}}
to visually inspect the detections of these candidates. We assigned a
class (class 1, good candidate) to detections with clear
extended X-ray emission and a class (class 2, weak candidate) to 
sources with faint and unclear emission. Classes 1 and 2 have 24 and 16 candidate objects, respectively.

Another classification was made based on the appearance of cluster
galaxies in the SDSS colour image. A very rough estimate of the
candidate distance was assigned to each source as follows: low
redshift ($z < 0.6$) and high redshift ($z > 0.6$). The candidates
expected at low and high redshifts are 15 and 25 sources,
respectively. Using the SDSS deep data in the Stripe 82, we expect to
detect clusters at $z \le 0.9$.

Table~\ref{tab:cand-cat} lists the sample of the remaining cluster candidates
(40 systems) with their X-ray positions, fluxes in the energy band
[0.5 - 2.0]~keV, and classifications (good or weak candidate and nearby 
or distant system). The column names of the table are listed
in  Table B.2 of Appendix B .



\section{Discovery of two merging galaxy cluster candidates}

During the visual inspection of the X-ray and SDSS colour images of
the cluster sample in our survey, we noted the small angular
separation between two pairs of extended X-ray sources with X-ray
emission between them. In addition, their optical counterparts have
similar redshifts, as reported in various projects. This drew our
attention to the cluster merger candidates in our cluster sample, therefore
we searched for cluster pairs with a physical separation smaller than
$R_{200}$ and a redshift difference of $\Delta z < 0.05$.  As a
result, we found only those two pairs that were spotted by eye during 
the visual inspection process.   
The first pair (3XMM J010606.7+004925 and 3XMM
J010610.0+005108) is at a redshift $z \sim$ 0.26, while the second pair
(3XMM J030617.3-000836 and 3XMM J030633.1-000350) is at 
$z \sim$ 0.11, the latter probably being two subclusters of Abell 0412
(A412 hereafter). We describe here the X-ray properties of the two
systems and build galaxy density maps to draw the distribution of
cluster galaxies and to estimate the significance of the galaxy
overdensities detected in the cluster regions.

\subsection{Twin galaxy clusters 3XMM J010606.7+004925 and 3XMM J010610.0+005108 at $z \sim 0.26$}

\begin{figure*}
 \centering
\includegraphics[width=6cm, viewport=14 10 399  395, clip]{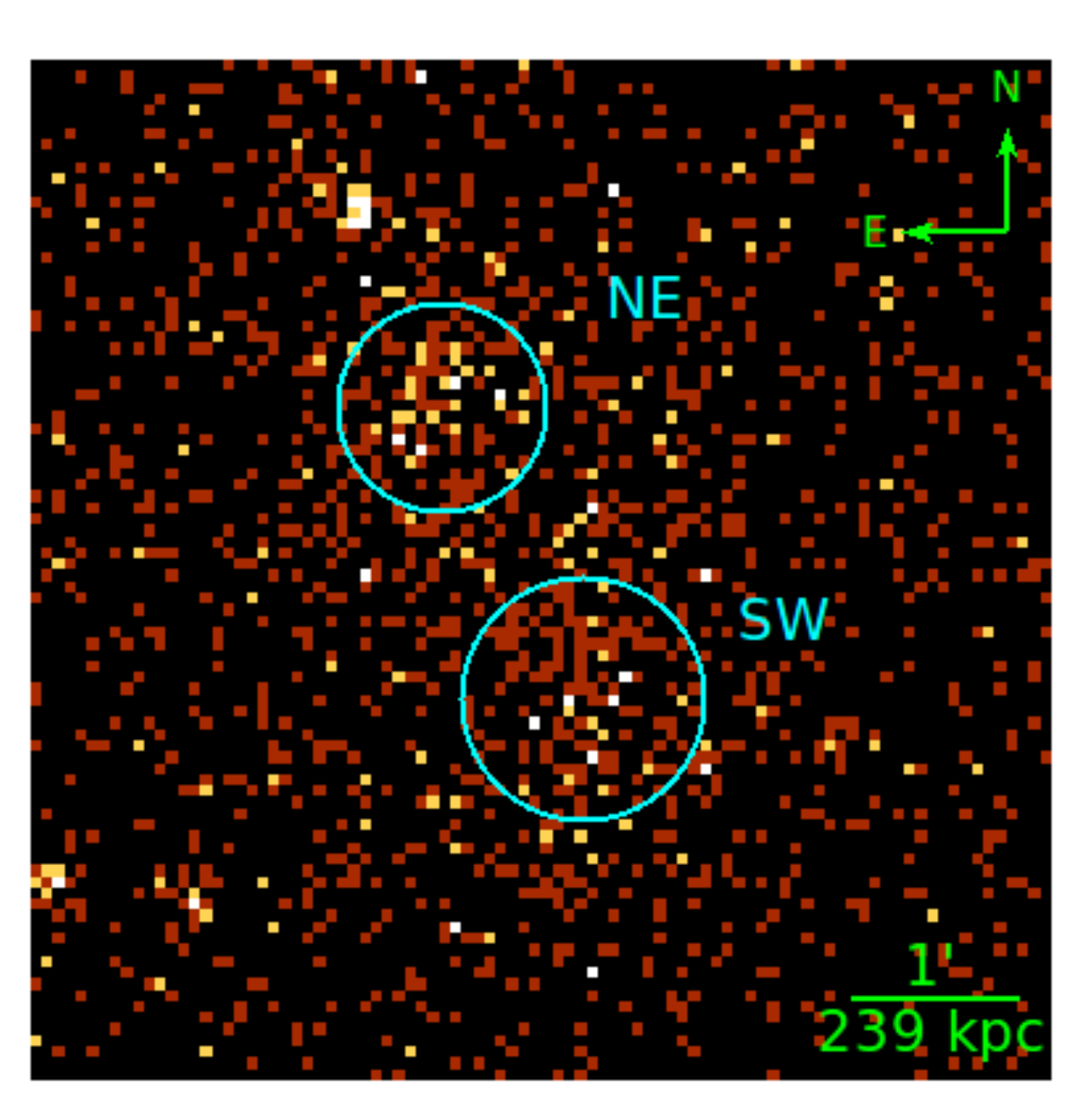}
\includegraphics[width=6cm, viewport=8  8 638 638, clip]{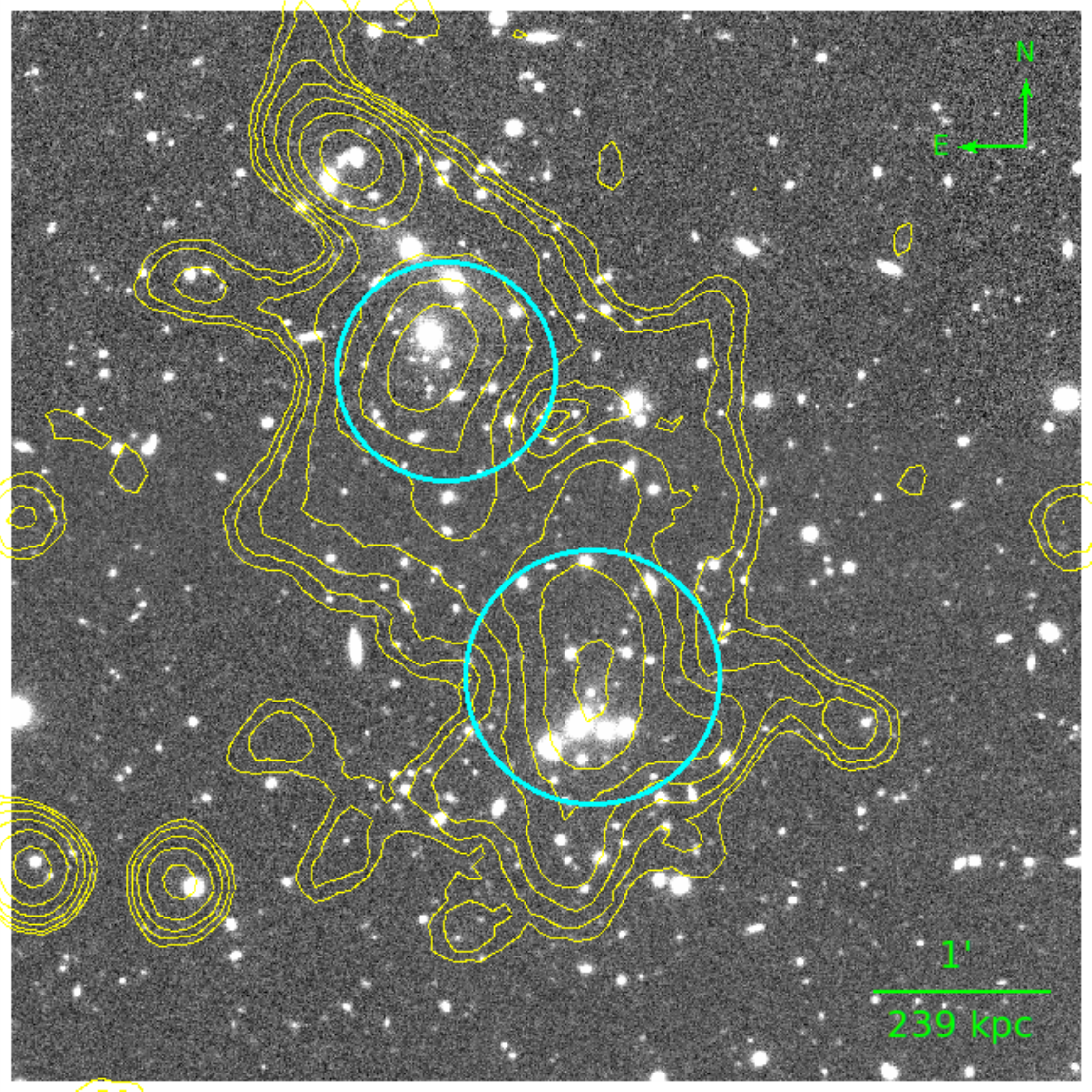}
\includegraphics[width=6cm, viewport=42 23 436 418, clip]{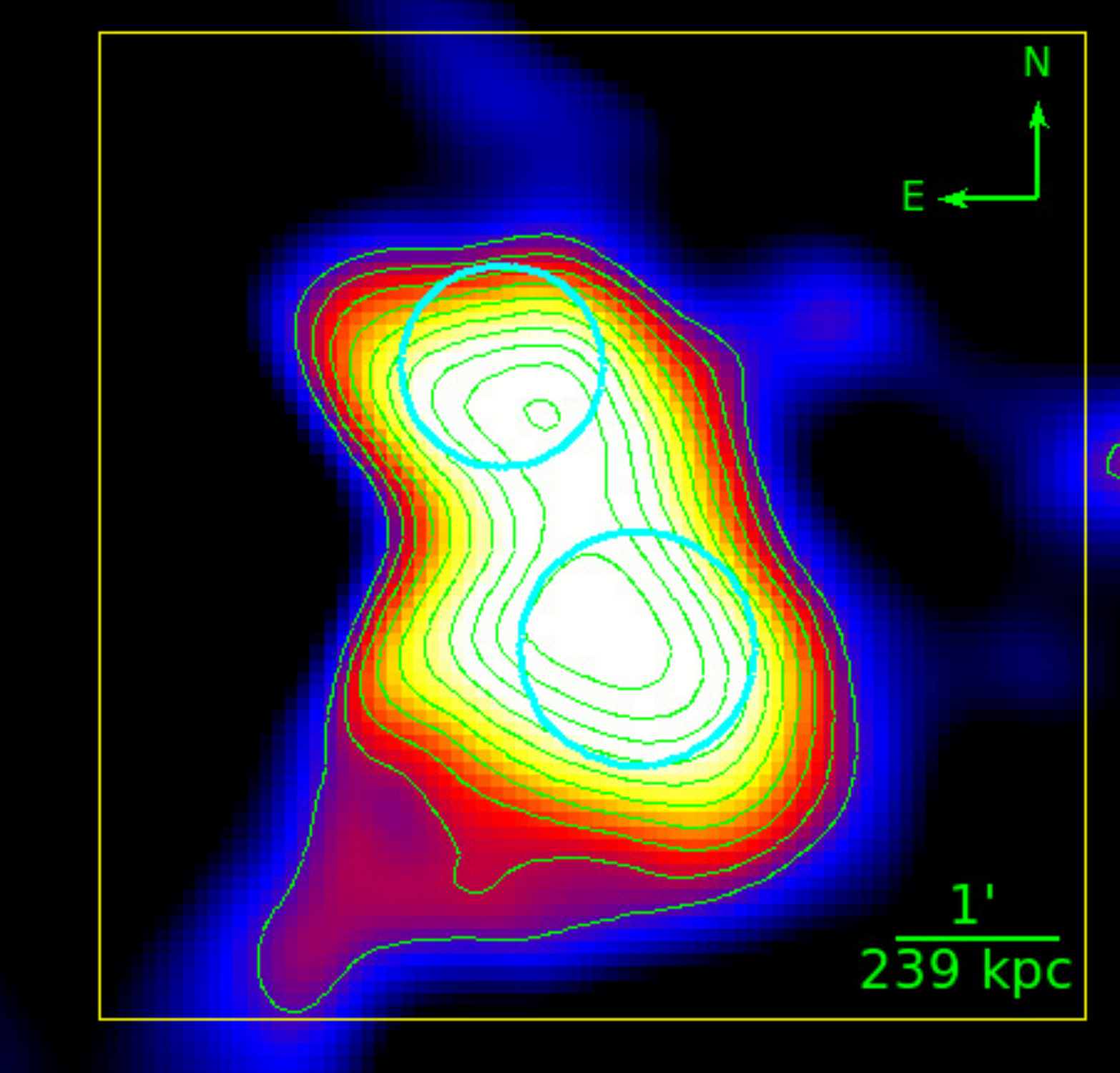}
\caption{Left: XMM-Newton EPIC (PN, MOS1, MOS2) combined image in
    the total energy band [0.2-12.0] keV for the twin galaxy clusters,   
    3XMM J010606.7+004925 (SW) and 3XMM J010610.0+005108 (NE).
    This observation, ObsID = 0150870201, is obtained as a reprocessed 
    product from the XMM-Newton Science Archive. Middle: deep SDSS image 
    in the r band from the Stripe 82 data corresponding to the X-ray 
    field. The overlaid X-ray surface brightness contours (yellow) 
    are in the total energy band [0.2-12.0] keV. 
    Right: Density map of the galaxies with photometric redshifts within
    $z_{CLG} \pm 0.03(1+z_{CLG})$ of the NE cluster redshift (see text). 
    The contours start at $3\sigma$ and increase by 1$\sigma$. 
    In the three images, the circles have radii equal to the cluster 
    core radii (SW: $r_c$ = 42.9 arcsec, NE: $r_c$ = 36.8 arcsec) and are
    centred on the X-ray emission peaks. The sizes of all the images are 
    $6\times 6$ arcmin$^2$ ($1.4\times1.4$ Mpc$^2$) centred on a middle 
    point between the two clusters.
    }
 \label{f:pair1_image}
\end{figure*}

\begin{figure*}
 \centering
\includegraphics[width=8cm]{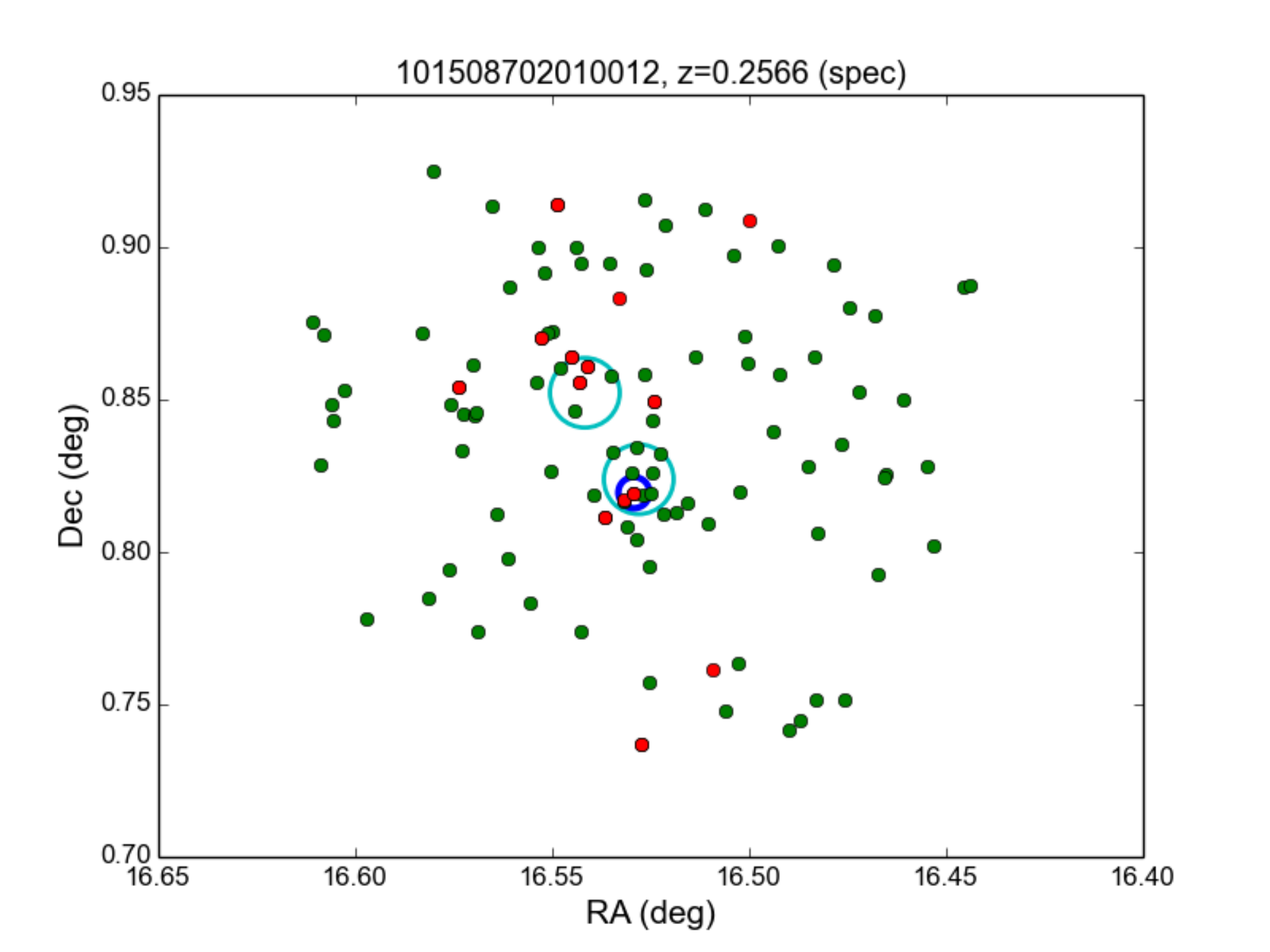}
\hspace{0.5cm}   
\includegraphics[width=7.5cm]{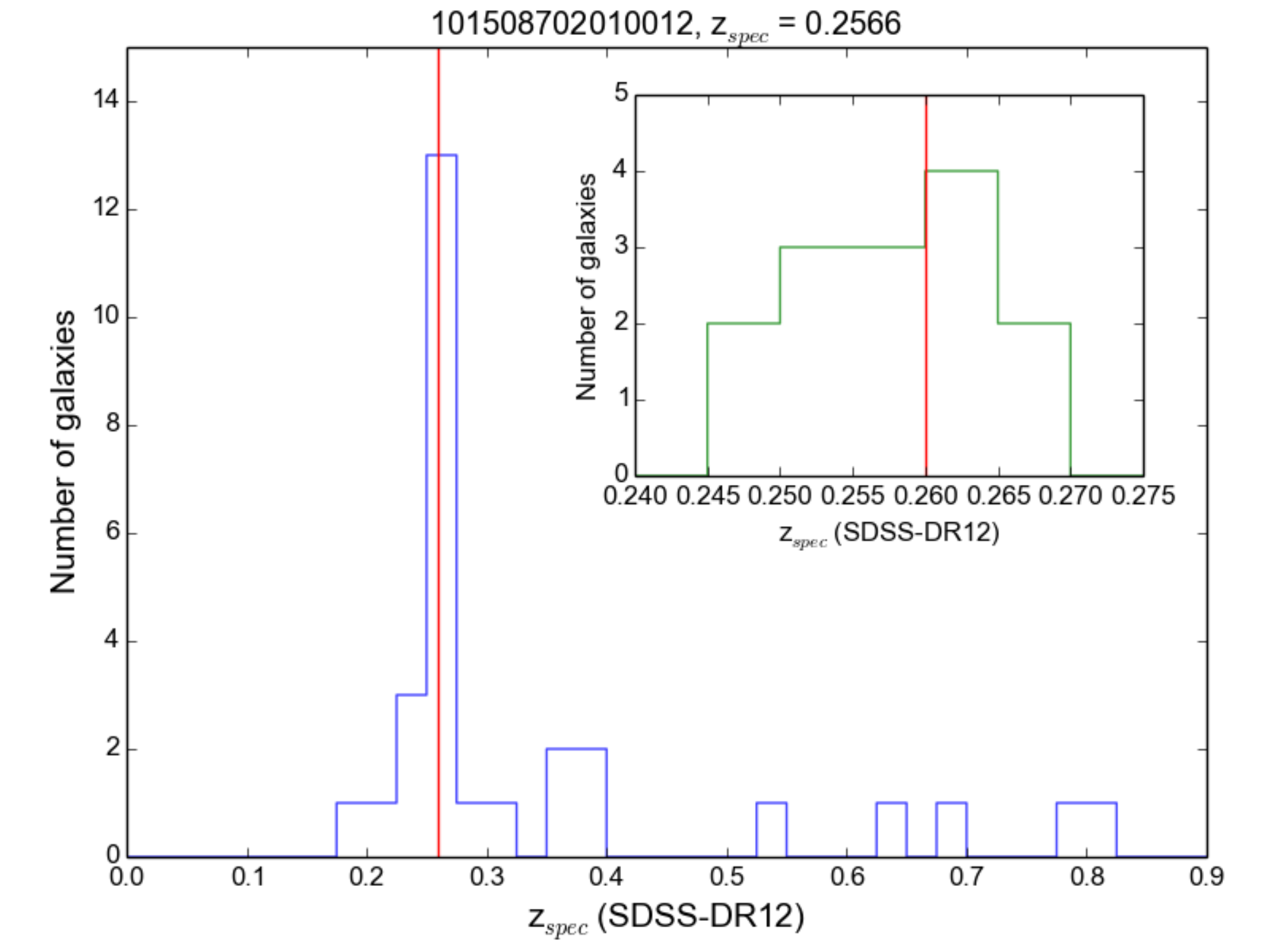}
\caption{Left: Sky distribution of the cluster galaxies selected based on their photometric redshifts (green dots, 99 galaxies) or spectroscopic redshifts (red dots overplotted on the green dots, 14 galaxies) within 1.5 Mpc (6.3 arcmin) centred on a position between the two systems, 3XMM J010606.7+004925 (SW) and 3XMM J010610.0+005108 (NE). The X-ray cluster positions of the two components are marked by the cyan circle centres, while the BCG is marked by the blue circle centre within the SW cluster. Right: Histogram of all galaxies with spectroscopic redshifts (29 galaxies, main figure) and of cluster galaxy members (14 galaxies, inset) within the aperture considered. In both histograms, the vertical red lines indicate the  redshift of the NE cluster.     
}
 \label{f:pair1_gal}
\end{figure*}

\addtocounter{table}{+2}
\begin{table*}
\small
\caption{\label{tab:pair1} X-ray parameters of the twin clusters 
based on available Chandra and XMM-Newton data. The columns are IAU name,
redshift, observation identifications, exposure times, core radii, 
X-ray photon counts, fluxes, luminosities, and masses.
}
  \begin{tabular}{c c c c c c c c c c }
  \hline\hline
  ID & Name    &  z & ObsID &  EXPT & $r_c$  & Counts & $F_{X}$  & $L_{X}$ & $M_{500}$ \\
     &         &    &       &  ks   & arcsec &      & $10^{-14}$\ erg\ cm$^{-2}$\ s$^{-1}$& $10^{42}$\ erg\ s$^{-1}$  & $10^{13}$\ M$_\odot$ \\ 
  \hline
  \multicolumn{10}{c}{Chandra observation \citep{Barkhouse06}}\\
  \hline
  SW & CXOMP J010607.0+004943 & 0.2767 & 2180 & 3.757 & 40.34  & 186.98 & 23.2 & 55.59  & -  \\
  NE & CXOMP J010610.3+005126 & 0.2630 & 2180 & 3.757 & 41.15  & 120.07 & 22.9 & 48.93  & -  \\
  \hline
   \multicolumn{10}{c}{XMM-Newton observation (This work)}\\
  \hline
  SW & 3XMM J010606.7+004925 & 0.2564 & 0150870201 & 2.392 & 42.9 & 359.5 & 13.8 & 62.68 & 11.84 \\
  NE & 3XMM J010610.0+005108 & 0.2566 & 0150870201 & 2.392 & 36.8 & 355.7 & 13.1 & 59.71 & 11.56 \\
   \hline
 \end{tabular}
\end{table*}

The two galaxy clusters, J010606.7+004925 (south-west, SW system) and
J010610.0+005108 (north-east, NE cluster), were detected as two
independent systems with slightly different redshift estimates in two
X-ray cluster surveys by \citet{Barkhouse06} and
\citet{Takey13}. Other galaxy cluster detection algorithms identified
the two clusters as a single system since they are close to each other
and have almost the same redshift \citep[][]{Goto02, Bahcall03,
  Lopes04, Koester07, Wen09, Hao10, Mehrtens12}.

\citet{Barkhouse06} conducted a survey of serendipitous X-ray selected
clusters in the framework of the Chandra Multiwavelength Project
(ChaMP) and detected 
the pair of clusters as two extended serendipitous sources at an off-axis
angle of about 3 arcmin in an observation with a short exposure time
of 3.8 ks. The redshift of the first system (SW), 0.2767, was obtained
from a possible association with an optical cluster from NED, while the
redshift for the second system (NE), 0.2630, was obtained from a galaxy
close to the X-ray emission peak.  Table~\ref{tab:pair1} lists the
estimated parameters of the two clusters based on these Chandra data.  
Barkhouse et al. did not state that they might form a merging system.

In the 2XMMi/SDSS galaxy cluster survey, these two clusters were
identified in X-rays \citep{Takey13}. Both systems were detected at
off-axis angles of $\sim$ 3.2 arcmin in the same XMM-Newton field. 
Their optical counterparts were identified and their redshifts, $z
\sim 0.26$, were estimated based on the few galaxies with available
spectroscopic redshifts in the SDSS-DR8, see the redshift values
listed in Table~\ref{tab:pair1}.

This pair of clusters was re-identified in our present cluster
survey. Figure~\ref{f:pair1_image} (left) shows the combined XMM-Newton
EPIC (PN, MOS1, MOS2) image of the cluster pair in the energy band
[0.2-12.0] keV, showing two separated extended sources. 
Table~\ref{tab:pair1} lists the main parameters derived for the two
clusters from the 3XMM-DR5 catalogue and our work (see the previous
section for the cluster luminosity and mass calculations). The
estimated X-ray luminosities and masses show that the two clusters have
similar properties (redshifts, core radii, fluxes, luminosities, and
masses), suggesting that they can be considered as twin galaxy clusters.

The angular separation between the two X-ray detections is 1.9 arcmin
(452 kpc, at the cluster redshift), which is smaller than $R_{500} =
2.9$ arcmin (685 kpc). The two clusters are marked in
Fig.~\ref{f:pair1_image} by circles centred on the X-ray positions
with radii equal to their core radii as given in the 3XMM-DR5
catalogue. The X-ray emission peaks coincide with two concentrations
of bright galaxies, as seen in the SDSS image in
Fig.~\ref{f:pair1_image} (middle). The X-ray contours clearly show two
separated extended sources with distortions in their X-ray surface
brightness distributions.  It is worth mentioning here that the
brightest cluster galaxies do not coincide with the peaks of
the innermost contours of both systems. Similar features have been
observed in other merging clusters \citep[e.g.][]{Maughan03}.

We note that the X-ray fluxes in the [0.5-2] keV band 
derived from Chandra and XMM-Newton observations are different, as
listed in Table~\ref{tab:pair1}. The  X-ray observations available from
Chandra (3.8 ks) and XMM-Newton (2.4 ks, good time interval) are too 
shallow to determine
the physical properties of the cluster pair with good accuracy. To
obtain a detailed analysis of the ICM and of the region between these
two clusters and to detect small-scale features such as bow shocks or
cold fronts, if present, deep X-ray observations are required. We have
applied to observe this cluster pair with Chandra. 

To show the distribution of cluster galaxies on the sky, we
selected all the galaxies within 1.5 Mpc (6.3 arcmin) from a middle point 
between the cluster pair with photometric redshifts in the interval $z_{CLG}\pm0.03(1+z_{CLG})$, where $z_{CLG}$ is the NE cluster redshift
0.2566. 
Cluster galaxies with spectroscopic redshifts in the range of 
$z_{CLG} \pm 0.01$ were also selected. Figure~\ref{f:pair1_gal} (left)
shows the sky distribution of cluster galaxies in the pair region
selected by their photometric (99 galaxies) and spectroscopic (14 objects) redshifts.
We can clearly see two concentrations of galaxies around the X-ray
centres, similar to those appearing in the SDSS r-band
image. Figure~\ref{f:pair1_gal} (right) presents the histogram of
spectroscopic  redshifts (29) in the chosen region. It shows a
peak at the cluster redshift, and the histogram zoomed around the cluster
redshift (14 galaxies) shows an almost Gaussian distribution.

The cluster galaxies selected based on their photometric redshifts were
used to compute a density map, with a fixed Gaussian kernel of
1.95~arcmin, a pixel size of
4.68~arcsec, and 10 bootstraps \citep{Martinet16}. To derive the
significance level of our detections, we estimated the mean background
and dispersion. For this, we drew the histogram of the pixel
intensities.  We applied a 2.5$\sigma$ clipping to eliminate the
pixels of the image that have high values and correspond to objects in
the image. We then redrew the histogram of the pixel intensities after
clipping and fit this distribution with a Gaussian. The mean value and
the width of the Gaussian will give the mean background
level and the dispersion, respectively, which we call $\sigma$. We then computed
the values of the contours corresponding to 3$\sigma$ detections as
the background plus 3$\sigma$. 

The galaxy density map (Fig.~\ref{f:pair1_image}, right) shows two
structures detected at 12$\sigma$ and coinciding with the X-ray
emission peaks. This confirms the two structures seen in X-ray and
optical images (Fig.~\ref{f:pair1_image}, left and middle). In
addition to the two clusters, we detected at 3$\sigma$ a galaxy group
south-west of the SW cluster (outside the field of 
Fig.~\ref{f:pair1_image}). The redshift of this group ($\sim$~0.25) is
almost the same as for the twin clusters, and it is located along the
same direction as the two cluster centres. The coordinates and the
spectroscopic redshift of the brightest galaxy in the group are
(16.483576, 0.714648) and $z=0.2515$. The brightest galaxy position
matches a cluster position (GMBCG J016.48356+00.71464 at a photometric
redshift of 0.251) identified by \citet{Hao10}. The separation between
the SW cluster and the brightest group galaxy is 7.1~arcmin (1.7
Mpc). This suggests that the cluster pair and the galaxy group are
located along a large-scale structure filament.  We searched for X-ray
sources around the galaxy group position but detected neither point
nor extended sources in the current XMM-Newton observation.


\subsection{Abell 0412 ($z \sim 0.11$)}

\begin{figure*}
\centering
 \includegraphics[width=6 cm, viewport= 16 16 554  554,clip]{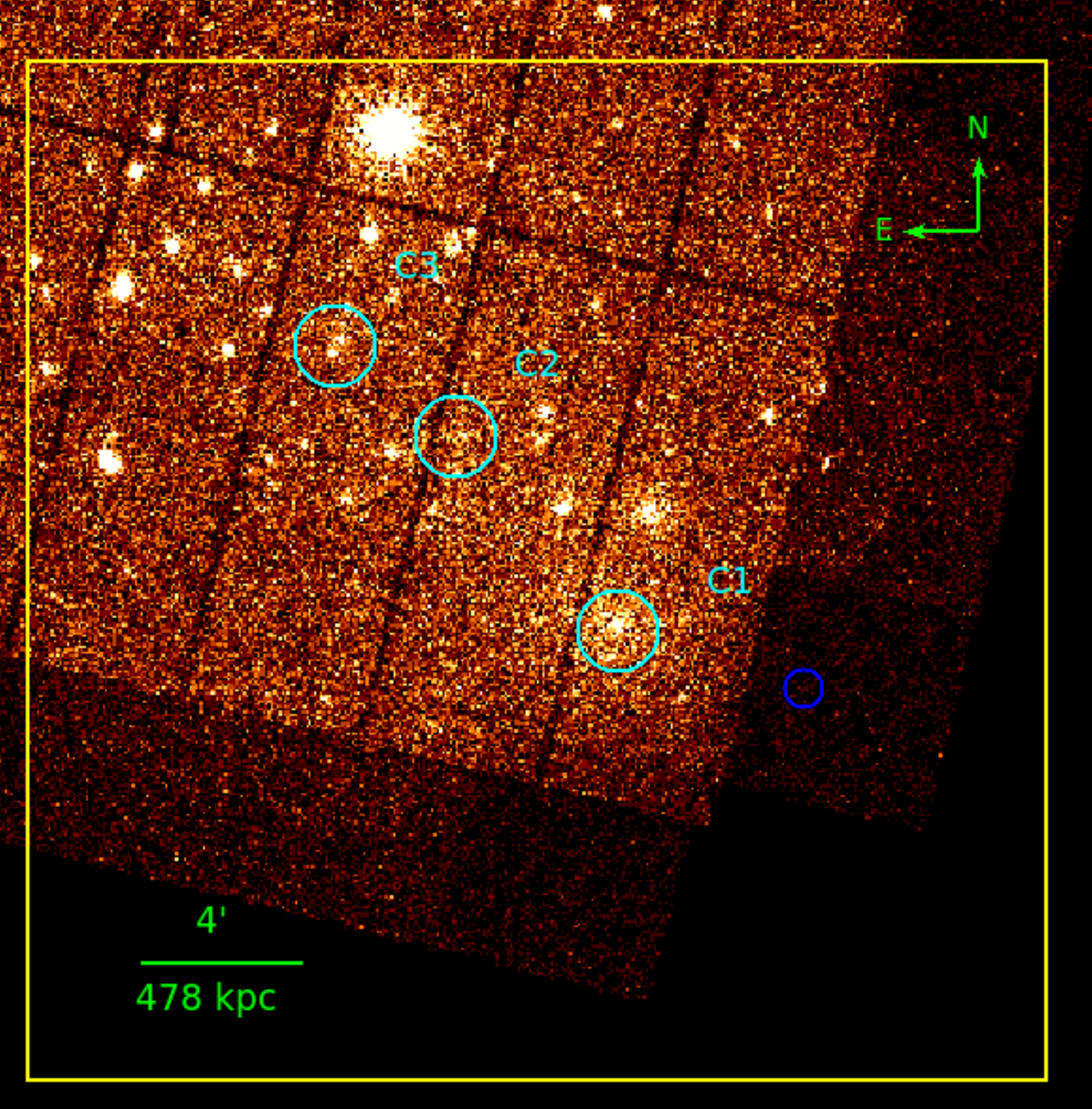}
 \includegraphics[width=6 cm, viewport= 14 18 622 626,clip]{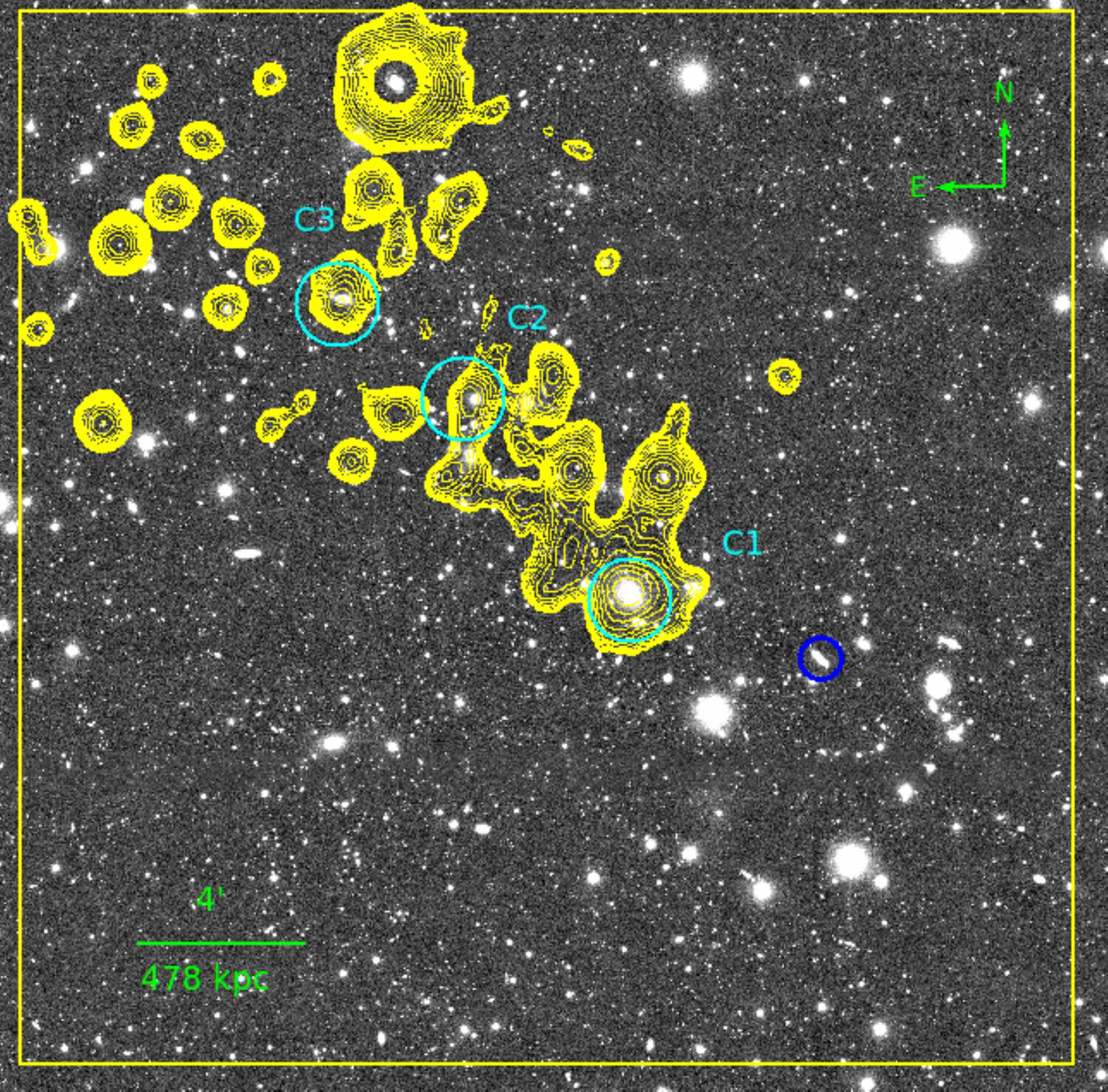}
 \includegraphics[width=6 cm, viewport= 16 10 398 392,clip]{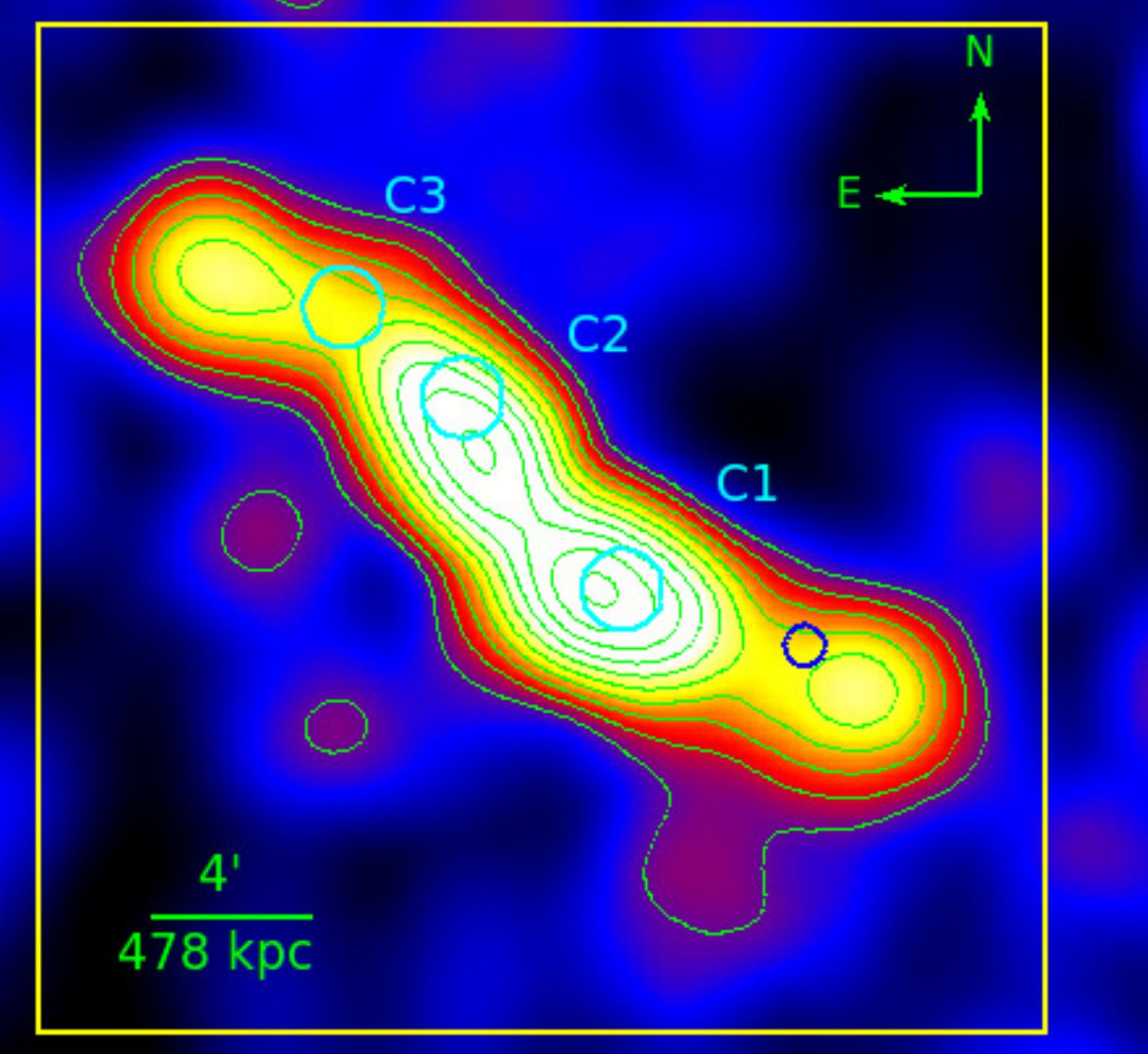}
 \caption{Left: XMM-Newton EPIC(PN, MOS1, MOS2) image in [0.2-12] keV
   of ObsID=0142610101 for Abell 0412 (C1) and the other two clusters 
   (C2 and C3). The blue circle indicates the BCG position, which 
   is located to the south-west of C1 position. 
   Middle: deep S82 image in the r band corresponding 
   to the X-ray image. The overplotted contours (yellow) are X-ray 
   surface brightness contours in [0.2-12] keV. Right:
   Density map of the galaxies with photometric redshifts within
   $z_{CLG} \pm 0.03(1+z_{CLG})$ of the cluster redshift (see
   text). The contours start at $4\sigma$ and increase by
   2$\sigma$. In the three images, the cyan circles have one 
   arcmin radii and are centred on the X-ray emission peaks of 
   the three (C1, C2, C3) X-ray sources. The sizes of all images are 
   $25\times 25$ arcmin$^2$ ($\sim 3\times3$ Mpc$^2$).
   }
 \label{f:A412_image}
\end{figure*}

\begin{figure*}
\centering
 \includegraphics[width=8.5 cm]{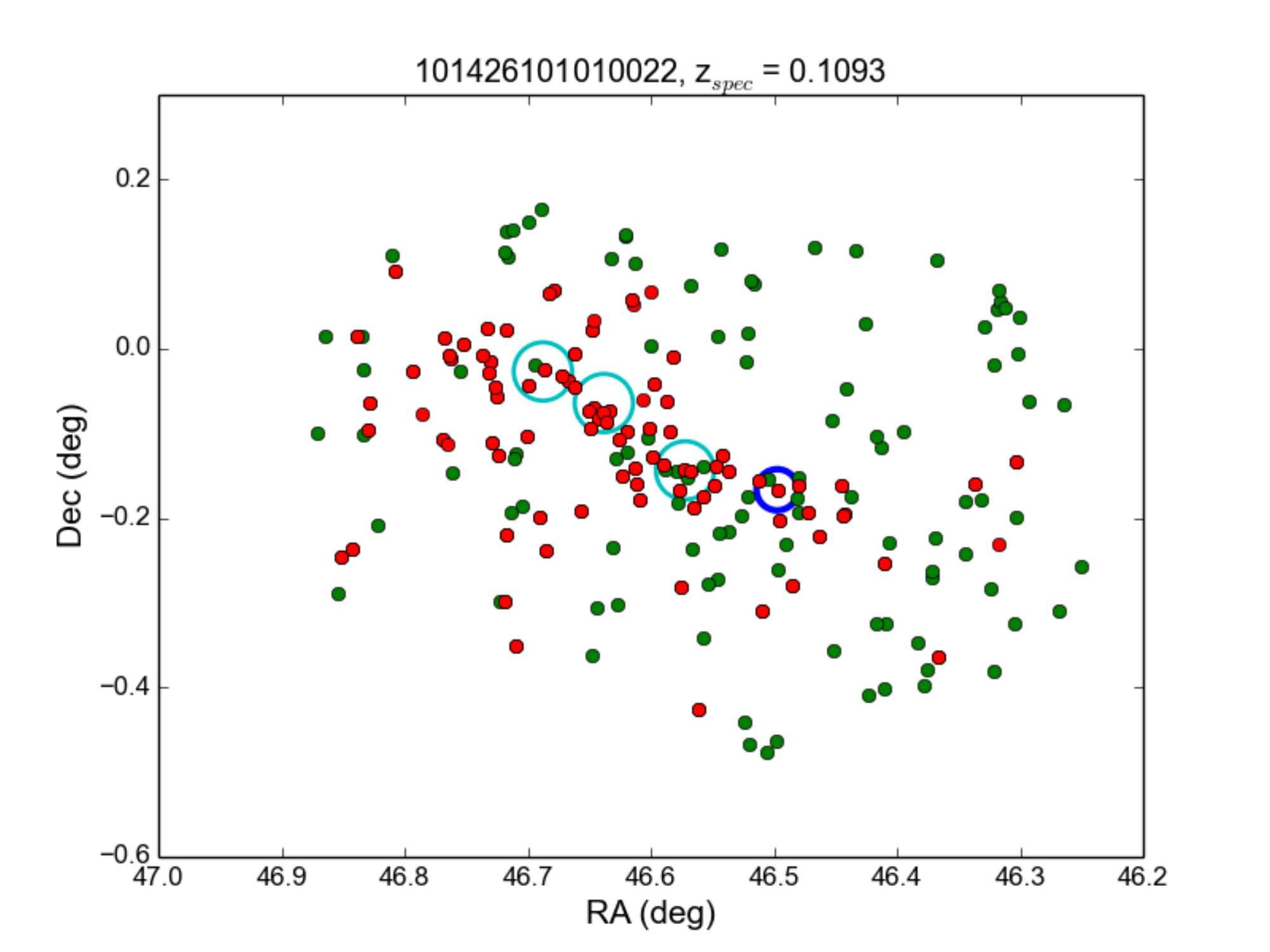}
 \hspace{0.5 cm}
 \includegraphics[width=8 cm]{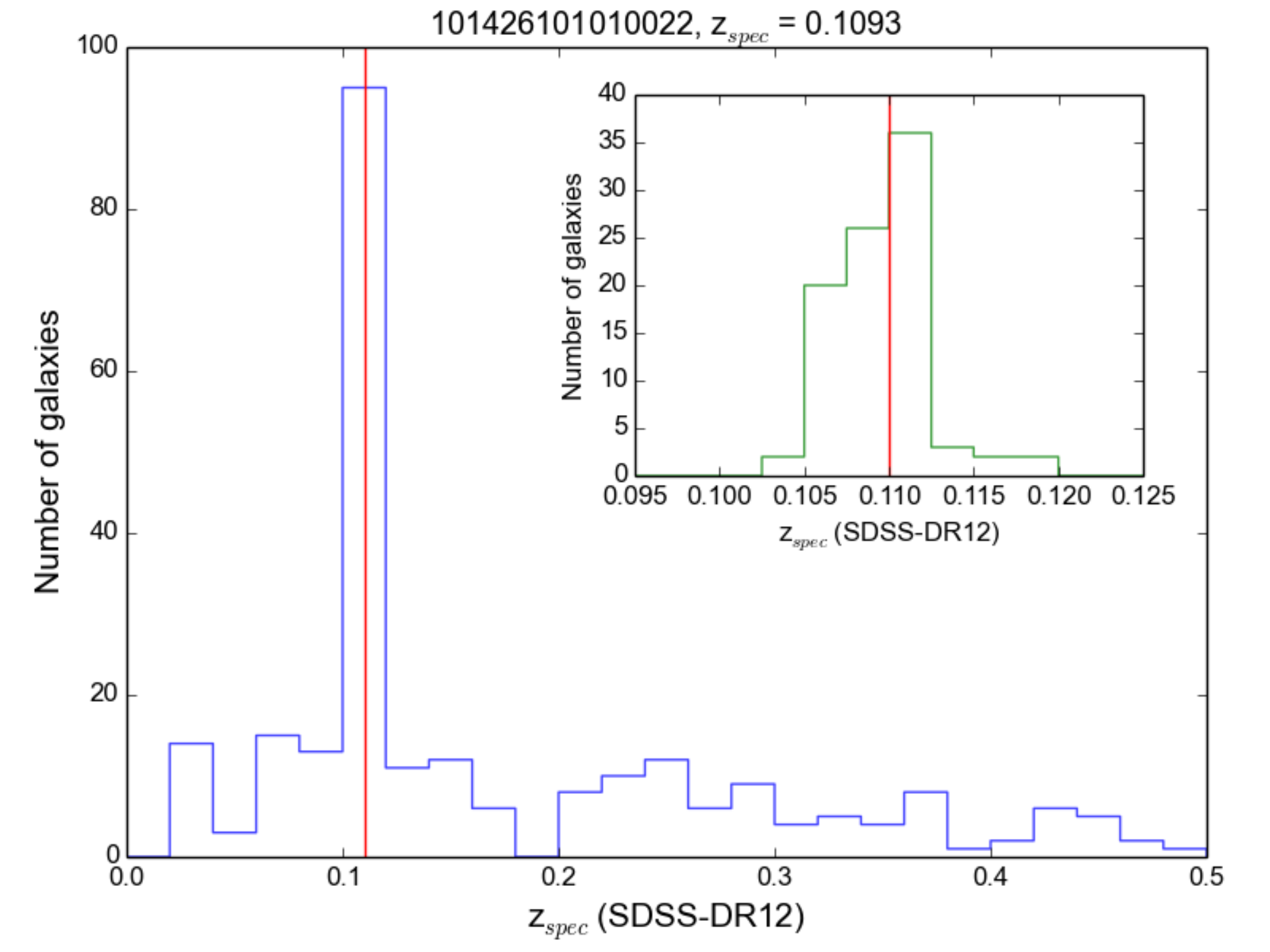}
 \hspace{0.5 cm}
 \caption{Left: Sky distribution of cluster galaxies within 2.5 Mpc
   (20.9 arcmin) from A412 selected based on their photometric (green
   dots, 191 galaxies) and spectroscopic (red dots overplotted on
   green dots, 91 galaxies) redshifts. The cyan circles are centred on
   the X-ray emission peaks of the three clusters, while the blue
   circle is centred on the BCG position. Right: spectroscopic
   redshift histogram of galaxies within 2.5 Mpc from A412. The inset
   shows a zoom around the mean cluster redshift. }
 \label{f:A412_gal}
\end{figure*}

\begin{table*}
\centering
\small 
 \caption{\label{A412} X-ray parameters for Abell 0412 (C1) and the other subclusters (C2 and C3) from the present work and the published ones. The columns are ID for each cluster, IAU Name, ra, dec, redshift, core radii (arcsec), X-ray photon counts, fluxes ($10^{-14}$\ erg\ cm$^{-2}$\ s$^{-1})$, luminosities ($10^{42}$\ erg\ s$^{-1}$), masses ($10^{13}$\ M$_\odot$), and references.  }
  \begin{tabular}{l l l l l l l l l l l}
  \hline\hline
  ID & IAU Name    & ra & dec & z &  $r_c$  & Counts & $F_{X}$  & $L_{500}$ & $M_{500}$ & References\\
  \hline
  C1 & 3XMM J030617.3-000836 & 46.57206 & -0.14361 & 0.1093 & 63.4 & 3281.4 & 13.2 & 10.0  & 5.34 & This work\\
    & XMMXCS J030617.2-000834.6 & 46.57167 & -0.14294 & 0.11 & -  & 2236.0 & - & 38 & - & \cite{Mehrtens12}\\
    & XMMU J030617.5+000827 & 46.57300 & -0.14100 & 0.109 & 217.2 & 3674.4 & - & - & - & \cite{Clerc12}\\
  \hline
  C2 & 3XMM J030633.1-000350 & 46.63804 & -0.06408 & 0.1235 & 21.3 & 414.7 & 1.0   & 1.1  & 1.8 & This work\\
     &XMMXCS J030634.0-000423.7 & 46.64167 & -0.07325 & 0.11 & - & 421.2 & - & 2.3 & -& \cite{Mehrtens12}\\
  \hline
  C3 & 2XMM J030645.0-000136 & 46.68755 & -0.02679 & 0.1178 & 10.3 & 536.6 & 0.65  & 2.1  & 2.5& \cite{Takey13}\\
     & XMMXCS J030644.2-000112.7 & 46.68417 & -0.02019 & 0.11 & - & 101.0 & - & - & - & \cite{Mehrtens12}\\
    \hline
 \end{tabular}
\end{table*}

A412 is a rich galaxy cluster identified by \citet{Abell58, Abell89}. 
It was later detected by various teams in SDSS data
by applying several algorithms \citep{Goto02, Bahcall03, Miller05,
  Koester07, Wen09, Hao10, Wen12, Smith12}. Combining X-ray data from
the ROSAT All-Sky Survey (RASS) and SDSS galaxies, A412 was also
identified by \citet{Schuecker04}. The redshift of the system reported
in these projects is $\sim 0.11$. It is worth mentioning that the
cluster centres in these catalogues can differ by as much as 5
arcmin ($\sim$ 600 kpc). \citet{Popesso07} searched for the X-ray
counterparts in the RASS of 137 optically selected Abell clusters in
the framework of the RASS-SDSS galaxy cluster survey. They found that
A412 is one of the 27 X-ray undetected clusters, which are probably in
the process of forming by undergoing a mass accretion phase.

Based on XMM-Newton observations, A412 is detected in three X-ray
cluster surveys (XCS, 2XMMi/SDSS, X-CLASS) with an offset from the
optical cluster centre of about 2 arcmin. In addition to the main
X-ray counterpart of A412, two other X-ray detected clusters
lie at a similar redshift in the same XMM-Newton field identified by
\citet{Takey11, Mehrtens12, Takey13}. The main X-ray cluster is
bright, while the two other systems are faint (see
Table~\ref{A412}). In these papers, they are not reported as
possible merging clusters.

In our current work, only two systems (A412 (C1, 3XMM J030617.3-000836) 
and C2, 3XMM J030633.1-000350) are re-detected as extended X-ray sources, 
while the third cluster (C3, 2XMM J030645.0-000136) is listed as 
a point-like source in the 3XMM catalogue. The XMM-Newton observation 
of the cluster field had an exposure time of $\sim$ 74 ks, but $\sim$ 40
percent of the exposure was lost through flares. There is no available
Chandra observation for this field to check the extent of C3. 
In our visual inspection, we noted the
similarity of their redshifts and the closeness of their
positions. Figure~\ref{f:A412_image} (left and middle) shows the X-ray
and optical images of the three galaxy clusters.  The separation
between C1 (A412) and the other two clusters is 6.2 arcmin (740 kpc)
for C2 and 9.9 arcmin (1177 kpc) for C3. The three systems are aligned
in the same direction, here again suggesting that we are detecting a
large-scale structure filament. Table~\ref{A412} lists the main
parameters for these clusters from our work and the mentioned
surveys. The X-ray luminosities from different surveys do not agree,
possibly because of the strong contamination by background flares and/or
different analysis algorithms.

We selected all cluster galaxies located within 2.5 Mpc (20.9 arcmin)
from the X-ray emission peak of the main component of A412 (C1). The
selection was based on a redshift interval of $z_{CLG} \pm
0.03(1+z_{CLG})$ (photometric) and $z_{CLG} \pm 0.01$ (spectroscopic),
where $z_{CLG}$ (0.1093) is the spectroscopic redshift of A412. In the
chosen aperture, 191 and 91 galaxies with photometric and
spectroscopic redshifts, respectively, lie in the cluster range. 

The spatial distribution of the selected cluster galaxies is presented in
Fig.~\ref{f:A412_gal} (left). It shows galaxy overdensities similar to 
those appearing in the SDSS image around X-ray positions. 
This means that the three cluster systems are
roughly in the plane of the sky. The brightest galaxy of the
selected galaxies can be considered as the BCG and is offset by
4.7 arcmin (561 kpc) from the main X-ray emission peak of
A412. Figure~\ref{f:A412_gal} (right) shows the spectroscopic redshift
histogram of all the galaxies within the chosen aperture (main figure)
and of cluster galaxies (inset). It shows a significant peak at the
cluster redshift and an almost Gaussian distribution of the
spectroscopic redshifts of cluster galaxies.

We built the galaxy density map for the photometrically selected
cluster galaxies in a similar way as for the other cluster pair. The
Gaussian kernel was fixed to 3.9~arcmin and the significance levels of
the detections were generated as described
above. Figure~\ref{f:A412_image} (right) shows the density map for the
cluster galaxies, with C1 and C2 detected at a high
significance level ($22\sigma$ and $20\sigma,$ respectively). Two
fainter structures are detected at a $12\sigma$ significance level,
one north-east of C3 and one south-west of C1 around the BCG position.
The C3 structure itself is detected at a lower significance level and
may belong to the structure detected to its north-east.  We therefore
seem to be observing a chain of four clusters here, the two brightest
being C1 and C2, and the other two being less massive.



\section{Summary and outlook}

We have conducted a galaxy cluster survey based on extended detections
in the third XMM-Newton serendipitous source catalogue (3XMM-DR5)
located in the footprint of SDSS Stripe~82. The survey comprises 94
X-ray cluster candidates detected in 74 XMM-Newton observations that
cover a survey area of 11.25 deg$^2$. We presented the cluster sample
extracted from the cluster candidates that were previously known as
galaxy clusters in the literature with redshift measurements. The
redshifts were obtained by cross-matching our X-ray cluster candidate
list with cluster catalogues extracted from X-ray (XCS, 2XMMi/SDSS,
X-CLASS) and optical (WH15, redMaPPer, Geach) surveys or by searching
NED. This yielded a cluster sample of 54 galaxy groups or clusters in the
redshift range 0.05-1.2 with a median of 0.36. Of these 54 clusters,
45 have spectroscopic confirmations from the matched catalogues or
NED. We also used the spectroscopic redshifts from SDSS-DR12 to 
re-confirm the redshifts of the cluster sample. These data enabled 
us to re-confirm redshifts of 46 clusters and to give spectroscopic 
confirmations for six clusters with only photometric redshifts. About 
one third of the cluster sample are newly discovered clusters in 
X-rays.

For each system in the cluster sample, we determined the X-ray
luminosity in the [0.5-2.0]~keV band based on the flux given in the
3XMM-DR5 catalogue. The luminosity was converted into the bolometric
luminosity within $R_{500}$ based on a scaling relation derived in the
current work. The estimated bolometric luminosity was then used to
measure the cluster mass based on the scaling relation of \cite{Pratt09}. We also presented the remaining X-ray cluster candidates (40
systems) with flags indicating a rough estimate of their expected
redshifts according to SDSS images and their probability of being real
sources according to X-ray images. This list includes 25 objects
expected to be at high redshift $z > 0.6$.  We also reported the 
detection of two cluster pairs and discussed their properties based on
the available X-ray and SDSS data. These systems probably have more 
than two components and deserve to be studied in more detail with 
deeper X-ray exposures with Chandra and XMM-Newton.  

For the remaining cluster candidates without redshift measurement, we
plan to estimate the cluster photometric redshifts based on deep SDSS
(Stripe~82 co-add) data \citep{Annis14}, or other deep optical/NIR
archival data, or to follow-up the candidates with no deep archival
data. When we have the redshifts for the majority of the X-ray cluster
candidates, we will investigate the X-ray luminosity-temperature
relation for our cluster sample that includes systems with a broad
range of luminosity and redshift. In addition, the galaxy luminosity
functions as well as the morphological analysis of cluster galaxies
will be investigated. We will also use our X-ray selected sample to
improve the centre positions of the clusters detected by the AMACFI
cluster finder \citep{Durret15}.


 
\begin{acknowledgements}

This work is supported by the Egyptian Science and Technology Development Fund (STDF) and the French Institute in Egypt (IFE) in cooperation with the Institut d'Astrophysique de Paris (IAP), France. F. D. acknowledges long-term support from CNES. 
We thank Hugo Capelato and Florian Sarron for their contribution to
the software to create galaxy density maps and compute significance
contour levels. We also thank the referee for the useful comments that 
helped to improve the paper.

This research has made use of data obtained from the 3XMM XMM-Newton serendipitous source catalogue compiled by the 10 institutes of the XMM-Newton Survey Science Centre selected by ESA. This work is based on observations obtained with XMM-Newton, an ESA science mission with instruments and contributions directly funded by ESA Member States and the USA (NASA). This research has made use of the NASA/IPAC Extragalactic Database (NED) which is operated by the Jet Propulsion Laboratory, California Institute of Technology, under contract with the National Aeronautics and Space Administration (NASA).

Funding for SDSS-III has been provided by the Alfred P. Sloan 
Foundation, the Participating Institutions, the National Science Foundation,
and the U.S. Department of Energy. The SDSS-III web site is 
\url{http://www.sdss3.org/}. 
SDSS-III is managed by the Astrophysical Research 
Consortium for the Participating Institutions of the SDSS-III Collaboration 
including the University of Arizona, the Brazilian Participation Group, 
Brookhaven National Laboratory, University of Cambridge, University of 
Florida, the French Participation Group, the German Participation Group, 
the Instituto de Astrofisica de Canarias, the Michigan State/Notre 
Dame/JINA Participation Group, Johns Hopkins University, Lawrence Berkeley 
National Laboratory, Max Planck Institute for Astrophysics, New Mexico State 
University, New York University, Ohio State University, Pennsylvania State 
University, University of Portsmouth, Princeton University, the Spanish 
Participation Group, University of Tokyo, University of Utah, Vanderbilt 
University, University of Virginia, University of Washington, and Yale 
University.

\end{acknowledgements}



\begin{appendix}

\section{List of XMM-Newton fields in the survey (Table~\ref{tab:obsids-list})}

\begin{table}
 \caption{\label{tab:obsids-list} List of XMM-Newton observations that are included in the 3XMM/SDSS Stripe 82 galaxy cluster survey. }
 \centering
 {\scriptsize 
 \begin{tabular}{l c c c c}
  \hline\hline
  OBSID    &      RA   &    Dec    & Target \\
  \hline
0305751001 &   1.19633 &   0.10525 &            SDSS0004+00 \\
0403760301 &   2.76258 &   0.86150 &                 UVLG03 \\
0403760601 &   4.34363 &  -0.92503 &                 UVLG01 \\
0403760101 &   4.39304 &  -0.95431 &                 UVLG01 \\
0407030101 &   5.58117 &   0.26472 &               4C+00.02 \\
0403160101 &   7.34350 &  -0.19786 &       J002937.0-001218 \\
0203690101 &   9.83037 &   0.84803 &                  HCG07 \\
0303561301 &  10.82721 &   0.03172 &          Haro0040.8-00 \\
0090070201 &  10.85479 &   0.83783 &                 UM 269 \\
0303562201 &  10.87621 &   0.00194 &          Haro0040.8-00 \\
0303110401 &  14.06721 &   0.55969 &        SDSS J0056+0032 \\
0150870201 &  16.55667 &   0.82206 &           BRI0103+0032 \\
0605391101 &  20.11646 &  -0.09786 &                    FG7 \\
0145450201 &  28.90667 &   0.48389 &        SDSS015543+0028 \\
0303110101 &  29.29692 &  -0.86883 &        SDSS J0157-0053 \\
0101640201 &  29.93642 &   0.41289 &               Mrk 1014 \\
0201090201 &  31.59017 &  -0.30761 &               MKN 1018 \\
0200730601 &  32.70213 &  -0.32494 &        SDSS J0210-0018 \\
0652400501 &  36.71258 &   0.60531 &                  DEEP2 \\
0652400601 &  37.11837 &   0.57278 &                  DEEP2 \\
0652400701 &  37.49567 &   0.56567 &                  DEEP2 \\
0652400801 &  37.81963 &   0.61531 &                  DEEP2 \\
0655380601 &  38.23179 &   0.43714 &          MCG+00-07-041 \\
0556213601 &  38.89858 &   0.19639 &    SDSS023540.90+00103 \\
0312190401 &  43.82312 &  -0.20111 &         NGC 1144 SWIFT \\
0606431301 &  44.79583 &   0.17175 &              0256-0000 \\
0041170101 &  45.68167 &   0.10825 &                 CFRS3h \\
0201120101 &  46.64783 &  -0.20856 &                   S2F5 \\
0142610101 &  46.69217 &   0.00097 &                  S2F1a \\
0402320201 &  53.64487 &   0.08525 &        SDSS J0334+0006 \\
0134920901 &  58.44812 &  -0.10111 &                SA95-42 \\
0203230501 &  58.69825 &   0.59892 &            Z03521+0028 \\
0304801201 & 323.39374 &  -0.84217 &               NGC 7089 \\
0655346843 & 332.31525 &  -0.96314 &              CFHTLS-W4 \\
0655346840 & 332.94739 &  -0.05908 &              CFHTLS-W4 \\
0655346842 & 333.59445 &  -0.97025 &              CFHTLS-W4 \\
0655346839 & 333.70404 &   0.81522 &              CFHTLS-W4 \\
0673000137 & 333.71628 &   0.14469 &        Stripe 82 X - 1 \\
0673000138 & 333.71707 &  -0.10533 &        Stripe 82 X - 1 \\
0673000139 & 333.71713 &  -0.35500 &        Stripe 82 X - 1 \\
0673000136 & 333.71716 &   0.39456 &        Stripe 82 X - 1 \\
0655346841 & 333.95367 &  -0.61422 &              CFHTLS-W4 \\
0673000141 & 333.96558 &   0.14436 &        Stripe 82 X - 1 \\
0673000140 & 333.96613 &  -0.35508 &        Stripe 82 X - 1 \\
0673000135 & 333.96683 &   0.64414 &        Stripe 82 X - 1 \\
0673000145 & 334.21521 &  -0.35517 &        Stripe 82 X - 1 \\
0673000134 & 334.21524 &   0.64325 &        Stripe 82 X - 1 \\
0673000143 & 334.21567 &   0.14431 &        Stripe 82 X - 1 \\
0673000144 & 334.21579 &  -0.10481 &        Stripe 82 X - 1 \\
0673000142 & 334.21628 &   0.39397 &        Stripe 82 X - 1 \\
0094310101 & 334.41550 &   0.23408 &                  SSA22 \\
0094310201 & 334.41568 &   0.23297 &                  SSA22 \\
0673000133 & 334.46457 &   0.64347 &        Stripe 82 X - 1 \\
0673000146 & 334.46478 &  -0.35442 &        Stripe 82 X - 1 \\
0673000150 & 334.71341 &   0.14508 &        Stripe 82 X - 1 \\
0673000149 & 334.71347 &  -0.10444 &        Stripe 82 X - 1 \\
0673000147 & 334.71381 &  -0.35431 &        Stripe 82 X - 1 \\
0673000151 & 334.71387 &   0.39350 &        Stripe 82 X - 1 \\
0673000132 & 334.71439 &   0.64372 &        Stripe 82 X - 1 \\
0673000152 & 334.96249 &   0.14358 &        Stripe 82 X - 1 \\
0673000148 & 334.96274 &  -0.35428 &        Stripe 82 X - 1 \\
0673000131 & 334.96307 &   0.64317 &        Stripe 82 X - 1 \\
0670020201 & 335.45612 &  -0.89961 &       SL2S J22214-0053 \\
0066950301 & 349.54251 &   0.28469 &               NGC 7589 \\
0652400901 & 351.61224 &   0.10564 &                  DEEP2 \\
0652401001 & 351.97433 &   0.14617 &                  DEEP2 \\
0673002301 & 352.21255 &  -0.31800 &        Stripe 82 X - 1 \\
0652401101 & 352.32999 &   0.16217 &                  DEEP2 \\
0652401201 & 352.63690 &   0.16261 &                  DEEP2 \\
0652401401 & 353.04343 &   0.11322 &                  DEEP2 \\
0652401301 & 353.40225 &   0.11578 &                  DEEP2 \\
0147580401 & 356.87891 &   0.88344 &       1AXGJ234725+0053 \\
0303110801 & 359.55432 &  -0.14042 &        SDSS J2358-0009 \\
0303110301 & 359.60184 &  -0.17114 &        SDSS J2358-0009 \\
  \hline
 \end{tabular}
 }
\end{table}



\section{Column description of cluster and cluster candidate catalogues (Table~\ref{tab:CLG-cat-cols} and Table~\ref{tab:cand-cat-cols} )}

\begin{table}
 \caption{\label{tab:CLG-cat-cols} List of column names of the cluster catalogue constructed from our survey. It consists of 54 galaxy groups and clusters with redshift measurements as well as X-ray properties (X-ray luminosity and mass).}
 \centering
 {\scriptsize
 \begin{tabular}{l l l }
  \hline\hline
  Cols       & Symbol &  Name (Units) \\
  \hline
  1  & DETID  &  X-ray detection number in the 3XMM-DR5 catalogue\\
  2  & IAU Name & IAU name of the X-ray source\\ 
  3  & RA     &  X-ray detection right ascension (J2000) (deg)\\
  4  & Dec    & X-ray detection declination (J2000) (deg)\\
  5  & OBSID  & XMM-Newton observation number\\
  6  & z      & Galaxy cluster redshift\\
  7  & scale  & Scale at the cluster redshift (kpc/$''$)\\
  8  & $F_{cat}$    & X-ray flux in [0.5-2.0] keV ($10^{-14}$\ erg\
                      cm$^{-2}$\ s$^{-1}$) \\
  9  & $eF_{cat}$   & Error in $F_{cat}$ in [0.5-2.0] keV ($10^{-14}$\ erg\
                      cm$^{-2}$\ s$^{-1}$)  \\
  10 & $L_{cat}$    & X-ray luminosity in [0.5-2.0] keV ($10^{42}$\ erg\
                      s$^{-1}$)\\
  11 & $eL_{cat}$   & Error in $L_{cat}$ in [0.5-2.0] keV ($10^{42}$\ erg\
                      s$^{-1}$)  \\
  12 & $R_{500}$    & Estimated physical radius (kpc)\\
  13 & $eR_{500}$   & Error in $R_{500}$ (kpc)\\
  14 & $L_{500}$    & X-ray bolometric luminosity within $R_{500}$ ($10^{42}$\ erg\ s$^{-1}$)\\
  15 & $eL_{500}$   & Error in $L_{500}$ ($10^{42}$\ erg\ s$^{-1}$)\\
  16 & $M_{500}$    & Cluster mass within $R_{500}$ ($10^{13}$\ M$_\odot$)\\
  17 & $eM_{500}$   & Error in $M_{500}$ ($10^{13}$\ M$_\odot$)\\
  18 & Nr. DETIDs   & Number of X-ray detections for the same source\\
  19 & X-ray cats   & Name of X-ray selected cluster catalogue; -9999 for the newly\\ 
     &              & discovered X-ray clusters based on XMM-Newton data.  \\
  20 & Optical cats & Name of optically selected cluster catalogue\\
  21 & z-type       & Redshift type (spec: spectroscopic, phot: photometric) \\
  22 & z-ref        & Reference to the chosen cluster redshift\\
  23 & $\bar z_{\rm s}$ & Mean redshift for cluster galaxies within $R_{200}$
                          with\\
     &                  & spectroscopic redshift from the SDSS-DR12 \\
  24 & $N_{z_{s}}$      & Number of cluster galaxies that are used to derive
                          $\bar z_{\rm s}$ \\
  \hline
 \end{tabular}
 }
\end{table}



\begin{table}
 \caption{\label{tab:cand-cat-cols} Columns of the catalogue of the remaining cluster candidates in our survey without redshift estimates.}
 \centering
 \scriptsize{
 \begin{tabular}{l l l }
  \hline\hline
  Cols       & Symbol &  Name (Units) \\
  \hline
  1  & DETID  & X-ray detection number in the 3XMM-DR5 catalogue\\
  2  & IAU Name & IAU name of the X-ray source\\
  3  & RA     & X-ray detection right ascension (J2000) (deg)\\
  4  & Dec    & X-ray detection declination (J2000) (deg)\\
  5  & OBSID  & XMM-Newton observation number \\
  6  & $F_{cat}$  & X-ray flux in [0.5-2.0] keV ($10^{-14}$\ erg\
                      cm$^{-2}$\ s$^{-1}$)\\
  7  & $eF_{cat}$ & Error in $F_{cat}$ in [0.5-2.0] keV ($10^{-14}$\ erg\
                      cm$^{-2}$\ s$^{-1}$)\\
  8  & Nr. DETIDs & Number of X-ray detections for the same source\\
  9  & class      & a flag of X-ray detection (1: good cluster candidate,\\ 
     &            & 2:weak cluster candidate)\\
  10 & redshift   & Expected redshift of the cluster candidate \\
     &            & (low redshift: $z < 0.6$, high redshift: $z > 0.6$)\\
  \hline
 \end{tabular}
 }
\end{table}

\end{appendix}



 \bibliographystyle{aa}
 \bibliography{refbib_thesis}
  


\clearpage

\onecolumn

\begin{landscape}

\addtocounter{table}{+1}

\begin{table}
\caption{\label{tab:CLG-cat} Cluster sample (54 objects) with redshift measurements in the literature. The X-ray parameters are determined based on the flux given in the 3XMM-DR5 catalogue.}
\end{table}

{\scriptsize
\begin{longtable}{c c c c c c c c c c c c c c c }
 \hline
 \hline
  \multicolumn{1}{c}{DETID\tablefootmark{a}} &
  \multicolumn{1}{c}{IAU Name\tablefootmark{a}} &
  \multicolumn{1}{c}{RA\tablefootmark{a}} &
  \multicolumn{1}{c}{Dec\tablefootmark{a}} &
  \multicolumn{1}{c}{OBSID\tablefootmark{a}} &
  \multicolumn{1}{c}{z\tablefootmark{b}} &
  \multicolumn{1}{c}{scale} &
  \multicolumn{1}{c}{$F_{cat}$\tablefootmark{a,c}} &
  \multicolumn{1}{c}{$\pm eF_{cat}$} &
  \multicolumn{1}{c}{$L_{cat}$\tablefootmark{d}} &
  \multicolumn{1}{c}{$\pm eL_{cat}$} &
  \multicolumn{1}{c}{$R_{500}$} &
  \multicolumn{1}{c}{$\pm eR_{500}$} &
  \multicolumn{1}{c}{$L_{500}$\tablefootmark{e}} &
  \multicolumn{1}{c}{$\pm eL_{500}$} \\

  \multicolumn{1}{c}{} &
  \multicolumn{1}{c}{} &
  \multicolumn{1}{c}{(deg)} &
  \multicolumn{1}{c}{(deg)} &
  \multicolumn{1}{c}{} &
  \multicolumn{1}{c}{} &
  \multicolumn{1}{c}{kpc/$''$} &
  \multicolumn{1}{c}{} &
  \multicolumn{1}{c}{} &
  \multicolumn{1}{c}{} &
  \multicolumn{1}{c}{} &
  \multicolumn{1}{c}{(kpc)} &
  \multicolumn{1}{c}{(kpc)} &
  \multicolumn{1}{c}{} &
  \multicolumn{1}{c}{}  \\

(1) &  (2)  &  (3)  & (4)  &   (5)   &  (6)  &   (7)  &   (8)  &  (9)  &  (10)  &  (11) &  (12) & (13) & (14) & (15) \\
\hline 
104037603010094 & 3XMM J001115.5+005152 &   2.81470 &   0.86462 & 0403760301 & 0.3622 & 5.05 &   1.00 & 0.14 &   4.45 &  0.62 &  486.50 & 35.42 &   10.94 &  1.45 \\
104037601010003 & 3XMM J001737.3-005240 &   4.40608 &  -0.87794 & 0403760101 & 0.2141 & 3.48 &  31.33 & 1.21 &  41.93 &  1.63 &  746.74 & 47.45 &   92.85 &  3.43 \\
104070301010041 & 3XMM J002223.3+001201 &   5.59736 &   0.20036 & 0407030101 & 0.2789 & 4.23 &   1.09 & 0.20 &   2.65 &  0.48 &  472.04 & 36.13 &    6.68 &  1.15 \\
104070301010056 & 3XMM J002314.4+001200 &   5.81017 &   0.20016 & 0407030101 & 0.2597 & 4.02 &   2.22 & 0.36 &   4.61 &  0.75 &  519.32 & 38.06 &   11.31 &  1.76 \\
104031601010005 & 3XMM J002928.6-001250 &   7.36918 &  -0.21393 & 0403160101 & 0.0600 & 1.16 &  14.30 & 0.73 &   1.24 &  0.06 &  472.80 & 35.02 &    3.22 &  0.16 \\
102036901010028 & 3XMM J003838.0+004351 &   9.65851 &   0.73108 & 0203690101 & 0.6955 & 7.13 &   4.88 & 0.40 & 104.35 &  8.54 &  641.90 & 41.28 &  221.49 & 17.28 \\
102036901010085 & 3XMM J003840.3+004747 &   9.66813 &   0.79659 & 0203690101 & 0.5553 & 6.44 &   1.47 & 0.18 &  18.22 &  2.26 &  536.42 & 36.71 &   41.94 &  4.97 \\
102036901010023 & 3XMM J003922.4+004809 &   9.84359 &   0.80277 & 0203690101 & 0.4145 & 5.49 &   5.31 & 0.25 &  32.59 &  1.52 &  639.13 & 41.14 &   73.01 &  3.24 \\
102036901010017 & 3XMM J003942.2+004533 &   9.92597 &   0.75926 & 0203690101 & 0.4156 & 5.50 &   2.43 & 0.18 &  15.01 &  1.11 &  567.36 & 37.83 &   34.87 &  2.45 \\
100900702010087 & 3XMM J004231.0+005112 &  10.62930 &   0.85336 & 0090070201 & 0.1579 & 2.73 &   5.60 & 0.97 &   3.81 &  0.66 &  533.89 & 39.52 &    9.44 &  1.56 \\
100900702010056 & 3XMM J004252.5+004300 &  10.71892 &   0.71692 & 0090070201 & 0.2697 & 4.13 &   5.86 & 0.53 &  13.23 &  1.19 &  606.63 & 40.61 &   30.90 &  2.65 \\
100900702010050 & 3XMM J004334.1+010107 &  10.89187 &   1.01811 & 0090070201 & 0.2000 & 3.30 &   8.71 & 0.73 &  10.01 &  0.84 &  604.65 & 40.67 &   23.69 &  1.89 \\
100900702010052 & 3XMM J004350.6+004731 &  10.96114 &   0.79216 & 0090070201 & 0.4754 & 5.94 &   5.12 & 0.44 &  43.59 &  3.71 &  643.84 & 41.83 &   96.34 &  7.82 \\
103035622010028 & 3XMM J004401.4+000644 &  11.00583 &   0.11226 & 0303562201 & 0.2187 & 3.54 &  10.47 & 1.34 &  14.70 &  1.88 &  634.60 & 43.13 &   34.18 &  4.18 \\
103031104010030 & 3XMM J005546.1+003839 &  13.94249 &   0.64422 & 0303110401 & 0.0665 & 1.28 &   3.52 & 0.53 &   0.38 &  0.06 &  393.06 & 32.78 &    1.04 &  0.15 \\
103031104010036 & 3XMM J005608.9+004106 &  14.03712 &   0.68510 & 0303110401 & 0.4607 & 5.84 &   2.95 & 0.66 &  23.25 &  5.24 &  590.17 & 43.31 &   52.91 & 11.38 \\
101508702010016 & 3XMM J010606.7+004925 &  16.52804 &   0.82388 & 0150870201 & 0.2564 & 3.98 &  13.80 & 2.43 &  27.77 &  4.90 &  684.62 & 47.46 &   62.68 & 10.54 \\
101508702010012 & 3XMM J010610.0+005108 &  16.54172 &   0.85226 & 0150870201 & 0.2566 & 3.99 &  13.09 & 1.84 &  26.40 &  3.70 &  679.24 & 45.82 &   59.71 &  7.98 \\
106053911010001 & 3XMM J012023.3-000444 &  20.09717 &  -0.07908 & 0605391101 & 0.0780 & 1.48 &  42.66 & 1.24 &   6.38 &  0.18 &  602.25 & 40.40 &   15.43 &  0.43 \\
101016402010005 & 3XMM J015917.1+003010 &  29.82144 &   0.50300 & 0101640201 & 0.3820 & 5.22 &  31.79 & 1.62 & 161.02 &  8.20 &  831.98 & 52.88 &  334.95 & 16.26 \\
101016402010022 & 3XMM J015953.1+001659 &  29.97148 &   0.28329 & 0101640201 & 0.7500 & 7.34 &   2.51 & 0.47 &  64.46 & 12.01 &  576.46 & 40.21 &  139.91 & 24.87 \\
101016402010018 & 3XMM J020019.2+001932 &  30.08002 &   0.32564 & 0101640201 & 0.6825 & 7.07 &   4.51 & 0.62 &  92.00 & 12.58 &  634.77 & 42.23 &  196.41 & 25.62 \\
102007306010070 & 3XMM J021012.6-001439 &  32.55263 &  -0.24443 & 0200730601 & 0.2828 & 4.27 &   2.21 & 0.41 &   5.57 &  1.04 &  527.52 & 39.09 &   13.55 &  2.42 \\
102007306010018 & 3XMM J021045.8-002156 &  32.69099 &  -0.36574 & 0200730601 & 0.3100 & 4.56 &   1.48 & 0.15 &   4.62 &  0.47 &  504.54 & 35.80 &   11.32 &  1.11 \\
106524006010012 & 3XMM J022825.8+003203 &  37.10780 &   0.53441 & 0652400601 & 0.3952 & 5.33 &   9.13 & 0.30 &  50.12 &  1.62 &  690.57 & 43.87 &  110.06 &  3.39 \\
106524006010017 & 3XMM J022830.5+003032 &  37.12738 &   0.50907 & 0652400601 & 0.7200 & 7.23 &   6.54 & 0.32 & 152.22 &  7.51 &  669.73 & 42.46 &  317.47 & 14.94 \\
106524007010008 & 3XMM J023026.7+003733 &  37.61157 &   0.62602 & 0652400701 & 0.8600 & 7.69 &   4.77 & 0.25 & 171.81 &  9.00 &  625.56 & 39.71 &  356.32 & 17.79 \\
106524008010044 & 3XMM J023058.5+004327 &  37.74395 &   0.72431 & 0652400801 & 0.4727 & 5.92 &   1.63 & 0.46 &  13.70 &  3.86 &  540.35 & 42.68 &   31.95 &  8.58 \\
106064313010011 & 3XMM J025846.5+001219 &  44.69388 &   0.20555 & 0606431301 & 0.2589 & 4.01 &   6.15 & 0.67 &  12.66 &  1.37 &  606.32 & 41.01 &   29.64 &  3.06 \\
106064313010004 & 3XMM J025932.5+001353 &  44.88574 &   0.23161 & 0606431301 & 0.1920 & 3.20 &  19.07 & 1.38 &  20.01 &  1.44 &  675.11 & 44.06 &   45.85 &  3.16 \\
100411701010097 & 3XMM J030145.7+000323 &  45.44072 &   0.05659 & 0041170101 & 0.6900 & 7.10 &   0.87 & 0.19 &  18.15 &  4.00 &  493.04 & 36.66 &   41.79 &  8.77 \\
100411701010033 & 3XMM J030205.6-000001 &  45.52347 &  -0.00053 & 0041170101 & 0.6500 & 6.93 &   2.33 & 0.21 &  42.15 &  3.85 &  574.92 & 37.73 &   93.31 &  8.12 \\
100411701010074 & 3XMM J030212.1-000132 &  45.55054 &  -0.02579 & 0041170101 & 1.1900 & 8.28 &   0.85 & 0.10 &  68.10 &  8.27 &  445.36 & 29.99 &  147.42 & 17.08 \\
100411701010113 & 3XMM J030212.1+001107 &  45.55082 &   0.18536 & 0041170101 & 0.6523 & 6.94 &   0.40 & 0.08 &   7.26 &  1.48 &  438.78 & 33.37 &   17.44 &  3.39 \\
100411701010112 & 3XMM J030317.5+001245 &  45.82296 &   0.21272 & 0041170101 & 0.5900 & 6.63 &   1.43 & 0.26 &  20.41 &  3.79 &  534.17 & 38.17 &   46.73 &  8.28 \\
101426101010024 & 3XMM J030614.1-000540 &  46.55923 &  -0.09474 & 0142610101 & 0.4249 & 5.57 &   3.23 & 0.20 &  21.01 &  1.29 &  593.91 & 38.94 &   48.04 &  2.80 \\
101426101010022 & 3XMM J030617.3-000836 &  46.57206 &  -0.14361 & 0142610101 & 0.1093 & 1.99 &  13.20 & 1.43 &   4.05 &  0.44 &  552.75 & 39.06 &    9.99 &  1.03 \\
101426101010059 & 3XMM J030633.1-000350 &  46.63804 &  -0.06408 & 0142610101 & 0.1235 & 2.22 &   1.00 & 0.14 &   0.40 &  0.05 &  385.13 & 31.95 &    1.10 &  0.14 \\
102011201010042 & 3XMM J030637.3-001801 &  46.65570 &  -0.30054 & 0201120101 & 0.4576 & 5.81 &   1.41 & 0.21 &  10.95 &  1.59 &  527.01 & 37.13 &   25.80 &  3.58 \\
104023202010027 & 3XMM J033446.2+001710 &  53.69279 &   0.28618 & 0402320201 & 0.3261 & 4.71 &   5.94 & 0.97 &  20.78 &  3.41 &  628.99 & 43.59 &   47.54 &  7.43 \\
101349209010028 & 3XMM J035416.9-001003 &  58.57060 &  -0.16751 & 0134920901 & 0.2100 & 3.43 &  14.95 & 2.81 &  19.17 &  3.60 &  664.05 & 46.86 &   44.01 &  7.89 \\
103048012010021 & 3XMM J213340.8-003841 & 323.41996 &  -0.64481 & 0304801201 & 0.2110 & 3.44 &  14.04 & 1.01 &  18.19 &  1.31 &  658.42 & 43.12 &   41.88 &  2.87 \\
106553468400021 & 3XMM J221211.0-000833 & 333.04618 &  -0.14275 & 0655346840 & 0.3643 & 5.07 &   5.18 & 0.96 &  23.47 &  4.35 &  626.45 & 44.11 &   53.38 &  9.43 \\
106553468390009 & 3XMM J221422.1+004712 & 333.59226 &   0.78680 & 0655346839 & 0.3202 & 4.66 &   4.23 & 0.66 &  14.17 &  2.20 &  595.31 & 41.49 &   33.00 &  4.89 \\
106553468390023 & 3XMM J221449.2+004707 & 333.70526 &   0.78535 & 0655346839 & 0.3171 & 4.63 &   3.81 & 0.66 &  12.50 &  2.16 &  585.10 & 41.51 &   29.29 &  4.83 \\
106730001440019 & 3XMM J221722.9-001013 & 334.34570 &  -0.17043 & 0673000144 & 0.3314 & 4.77 &   4.17 & 1.10 &  15.16 &  3.99 &  597.55 & 45.87 &   35.20 &  8.84 \\
106700202010013 & 3XMM J222144.0-005306 & 335.43347 &  -0.88513 & 0670020201 & 0.3363 & 4.81 &   4.13 & 0.52 &  15.52 &  1.95 &  597.94 & 40.71 &   35.98 &  4.30 \\
106524009010093 & 3XMM J232540.3+001447 & 351.41827 &   0.24660 & 0652400901 & 0.7900 & 7.48 &   0.21 & 0.16 &   6.12 &  4.78 &  392.40 & 54.44 &   14.82 & 11.03 \\
106524009010085 & 3XMM J232613.8+000706 & 351.55771 &   0.11844 & 0652400901 & 0.3844 & 5.24 &   0.58 & 0.11 &   2.96 &  0.56 &  451.04 & 34.84 &    7.41 &  1.35 \\
106524010010078 & 3XMM J232742.1+001406 & 351.92550 &   0.23522 & 0652401001 & 0.4441 & 5.72 &   0.81 & 0.16 &   5.85 &  1.17 &  482.78 & 36.45 &   14.18 &  2.71 \\
106524010010043 & 3XMM J232809.0+001116 & 352.03771 &   0.18778 & 0652401001 & 0.2780 & 4.22 &   3.12 & 0.46 &   7.54 &  1.11 &  554.07 & 39.27 &   18.09 &  2.53 \\
106524011010056 & 3XMM J232925.6+000554 & 352.35668 &   0.09849 & 0652401101 & 0.4021 & 5.39 &   0.75 & 0.11 &   4.31 &  0.62 &  472.69 & 34.70 &   10.61 &  1.46 \\
106524014010034 & 3XMM J233138.1+000738 & 352.90912 &   0.12725 & 0652401401 & 0.2344 & 3.73 &   2.53 & 0.18 &   4.15 &  0.29 &  518.44 & 36.30 &   10.23 &  0.68 \\
106524013010039 & 3XMM J233328.1-000123 & 353.36739 &  -0.02308 & 0652401301 & 0.4620 & 5.85 &   2.76 & 0.31 &  21.95 &  2.48 &  584.53 & 39.27 &   50.08 &  5.40 \\

\hline
\end{longtable}
}

 \end{landscape}

\clearpage
%

{\scriptsize 
\begin{longtable}{c  c c c c c c c c c c c c c }
\hline
\hline
  \multicolumn{1}{c}{DETID\tablefootmark{a}} &
  \multicolumn{1}{c}{$M_{500}$\tablefootmark{f}} &
  \multicolumn{1}{c}{$\pm eM_{500}$} &
  \multicolumn{1}{c}{Nr. \tablefootmark{a}} &
  \multicolumn{1}{c}{X-ray\tablefootmark{g}} &
  \multicolumn{1}{c}{Optical\tablefootmark{g}} &
  \multicolumn{1}{c}{z-type\tablefootmark{g}} &
  \multicolumn{1}{c}{z-ref\tablefootmark{g}} &
  \multicolumn{1}{c}{$\bar z_{\rm s}$\tablefootmark{h}} &
  \multicolumn{1}{c}{$N_{z_{s}}$\tablefootmark{h}} \\
  
  \multicolumn{1}{c}{}&
  \multicolumn{1}{c}{} &
  \multicolumn{1}{c}{} &
  \multicolumn{1}{c}{DETIDs} &
  \multicolumn{1}{c}{Catalogue} &
  \multicolumn{1}{c}{Catalogue} &
  \multicolumn{1}{c}{} &
  \multicolumn{1}{c}{} &
  \multicolumn{1}{c}{DR12} &
  \multicolumn{1}{c}{DR12} \\

  (1)  &  (16) & (17)  & (18)  &  (19)  &  (20)   &  (21) &  (22)  &   (23) &   (24) \\
\hline
104037603010094 &  4.78 &  1.04 & 1 &                 2XMM &                     Geach &          spec &                      2XMM & 0.3626 &  2 \\
104037601010003 & 14.67 &  2.80 & 3 &             2XMM-XCS &        WH15-redMaPPer-Geach &          spec &                  2XMM-XCS & 0.2056 &  9 \\
104070301010041 &  3.98 &  0.91 & 1 &                 2XMM &                      WH15 &          spec &                      2XMM & 0.2791 &  1 \\
104070301010056 &  5.19 &  1.14 & 1 &             2XMM-XCS &        WH15-redMaPPer-Geach &          spec &                  2XMM-XCS & 0.2584 &  3 \\
104031601010005 &  3.18 &  0.71 & 1 &                -9999 &                      WH15 &     spec &                      WH15 & 0.0603 & 54 \\
102036901010028 & 16.29 &  3.14 & 1 &                -9999 &                WH15-Geach &     spec &                WH15-Geach & 0.6954 &  2 \\
102036901010085 &  8.04 &  1.65 & 1 &                 2XMM &                     Geach &          spec &                      2XMM & 0.5555 &  5 \\
102036901010023 & 11.52 &  2.23 & 1 &          2XMM-XCLASS &                WH15-Geach &          spec &               2XMM-XCLASS & 0.4140 &  4 \\
102036901010017 &  8.07 &  1.61 & 1 &                 2XMM &                WH15-Geach &          spec &                      2XMM & 0.4151 &  5 \\
100900702010087 &  5.05 &  1.12 & 1 &             2XMM-XCS &                      WH15 &          spec &                  2XMM-XCS & 0.1544 &  5 \\
100900702010056 &  8.36 &  1.68 & 1 &             2XMM-XCS &        WH15-redMaPPer-Geach &          spec &                  2XMM-XCS & 0.2697 &  5 \\
100900702010050 &  7.67 &  1.55 & 2 &             XCS-2XMM &                WH15-Geach &     spec &                  XCS-2XMM & 0.1959 &  4 \\
100900702010052 & 12.65 &  2.47 & 1 &                 2XMM &            WH15-redMaPPer-Geach &          spec &                      2XMM & 0.4758 &  1 \\
103035622010028 &  9.05 &  1.85 & 1 &                 2XMM &            WH15-redMaPPer-Geach &          spec &                      2XMM & 0.2186 &  5 \\
103031104010030 &  1.84 &  0.46 & 1 &                  XCS &                      WH15 &     spec &                      WH15 & 0.0672 & 20 \\
103031104010036 &  9.57 &  2.11 & 1 &                 2XMM &                     -9999 &          phot &                      2XMM & -9999 & -9999 \\
101508702010016 & 11.84 &  2.46 & 1 &                 2XMM &                     -9999 &          spec &                      2XMM & 0.2580 & 11 \\
101508702010012 & 11.56 &  2.34 & 1 &                 2XMM &            WH15-redMaPPer &          spec &                      2XMM & 0.2582 & 14 \\
106053911010001 &  6.69 &  1.35 & 1 &                -9999 &                      WH15 &     spec &                      WH15 & 0.0801 & 19 \\
101016402010005 & 24.48 &  4.67 & 1 &                 2XMM &                      WH15 &           spec &                      2XMM & 0.3721 &  2 \\
101016402010022 & 12.59 &  2.64 & 1 &                  XCS &                     -9999 &            phot &                       XCS & -9999 & -9999 \\
101016402010018 & 15.51 &  3.10 & 1 &             2XMM-XCS &                     -9999 &          spec &                  2XMM-XCS & 0.6825 &  1 \\
102007306010070 &  5.58 &  1.24 & 1 &       2XMM-XCS-XCLASS &                     Geach &          spec &           2XMM-XCS-XCLASS & 0.2828 &  1 \\
102007306010018 &  5.03 &  1.07 & 1 &                  XCS &                      WH15 &     spec &                       XCS & 0.3161 &  2 \\
106524006010012 & 14.21 &  2.71 & 1 &                -9999 &         WH15-redMaPPer-Geach &     spec &      WH15-redMaPPer-Geach & 0.3875 &  3 \\
106524006010017 & 19.05 &  3.62 & 1 &                -9999 &                     -9999 &            phot &                       NED & 0.7214 &  2 \\
106524007010008 & 18.34 &  3.49 & 2 &                -9999 &                     -9999 &            spec &                       NED & -9999 & -9999 \\
106524008010044 &  7.45 &  1.77 & 1 &                -9999 &                WH15-Geach &     spec &                WH15-Geach & 0.4727 &  3 \\
106064313010011 &  8.25 &  1.67 & 1 &                -9999 &         WH15-redMaPPer-Geach &     spec &      WH15-redMaPPer-Geach & 0.2547 &  5 \\
106064313010004 & 10.59 &  2.07 & 1 &                -9999 &         WH15-redMaPPer-Geach &     spec &      WH15-redMaPPer-Geach & 0.1917 &  5 \\
100411701010097 &  7.33 &  1.64 & 1 &                  XCS &                     -9999 &     spec &                       XCS & -9999 & -9999 \\
100411701010033 & 11.09 &  2.18 & 1 &           XCS-XCLASS &                     -9999 &     spec &                XCS-XCLASS & -9999 & -9999 \\
100411701010074 &  9.69 &  1.96 & 1 &                  XCS &                     -9999 &     spec &                       XCS & -9999 & -9999 \\
100411701010113 &  4.94 &  1.13 & 1 &                 2XMM &                WH15-Geach &          spec &                      2XMM & 0.6523 &  1 \\
100411701010112 &  8.28 &  1.77 & 1 &                  XCS &                      WH15 &     spec &                       XCS & 0.6049 &  1 \\
101426101010024 &  9.36 &  1.84 & 2 &       2XMM-XCS-XCLASS &                     -9999 &          spec &           2XMM-XCS-XCLASS & 0.4249 &  1 \\
101426101010022 &  5.34 &  1.13 & 2 &       2XMM-XCS-XCLASS &                WH15-Geach &          spec &           2XMM-XCS-XCLASS & 0.1096 & 37 \\
101426101010059 &  1.83 &  0.46 & 1 &                 2XMM &                     -9999 &          spec &                      2XMM & 0.1122 & 23 \\
102011201010042 &  6.79 &  1.44 & 1 &             2XMM-XCS &                WH15-Geach &          spec &                  2XMM-XCS & 0.4576 &  1 \\
104023202010027 &  9.92 &  2.06 & 1 &             2XMM-XCS &                WH15-Geach &          spec &                  2XMM-XCS & 0.3270 &  2 \\
101349209010028 & 10.27 &  2.17 & 1 &             XCS-2XMM &                     -9999 &     spec &                  XCS-2XMM & 0.2143 &  2 \\
103048012010021 & 10.02 &  1.97 & 2 &               XCLASS &                     -9999 &     phot &                    XCLASS & -9999 & -9999 \\
106553468400021 & 10.24 &  2.16 & 1 &                -9999 &            WH15-redMaPPer &     spec &            WH15-redMaPPer & 0.3643 &  3 \\
106553468390009 &  8.36 &  1.75 & 1 &                -9999 &         WH15-redMaPPer-Geach &     spec &      WH15-redMaPPer-Geach & 0.3202 &  2 \\
106553468390023 &  7.91 &  1.68 & 1 &                -9999 &                WH15-Geach &     spec &                WH15-Geach & 0.3174 &  3 \\
106730001440019 &  8.56 &  1.97 & 1 &                -9999 &         WH15-redMaPPer-Geach &     spec &      WH15-redMaPPer-Geach & 0.3314 &  2 \\
106700202010013 &  8.62 &  1.76 & 1 &                -9999 &                     Geach &     phot &                     Geach & 0.3353 &  2 \\
106524009010093 &  4.17 &  1.73 & 1 &                -9999 &                     -9999 &            spec &                       NED & -9999 & -9999 \\
106524009010085 &  3.91 &  0.91 & 1 &                -9999 &                     Geach &     phot &                     Geach & 0.4261 &  1 \\
106524010010078 &  5.14 &  1.16 & 1 &                -9999 &                     Geach &     phot &                     Geach & 0.4451 &  1 \\
106524010010043 &  6.43 &  1.37 & 1 &                -9999 &                      WH15 &     spec &                      WH15 & 0.2780 &  3 \\
106524011010056 &  4.59 &  1.01 & 1 &                -9999 &                     -9999 &         spec &                       NED & 0.4019 &  1 \\
106524014010034 &  5.02 &  1.05 & 1 &                -9999 &                     Geach &     phot &                     Geach & 0.2238 &  1 \\
106524013010039 &  9.32 &  1.88 & 1 &                -9999 &                     Geach &     phot &                     Geach & 0.5120 &  2 \\

\hline
\end{longtable}
} \tablefoot{The full cluster catalogue is available online at the
  CDS.}  \tablefoottext{a}{Parameters for X-ray detections extracted
  from the 3XMM-DR5 catalogue.}  \tablefoottext{b}{Cluster redshift
  with type (spectroscopic or photometric redshift) as given in
  Col.~(21), obtained from the first catalogue in Col.~22.}
\tablefoottext{c}{X-ray flux, $F_{cat}$ [0.5 - 2.0]~keV, and its
  errors in units of $10^{-14}$\ erg\ cm$^{-2}$\ s$^{-1}$ .}
\tablefoottext{d}{X-ray luminosity, $L_{cat}$ [0.5 - 2.0]~keV, and its
  errors in units of $10^{42}$\ erg\ s$^{-1}$.}
\tablefoottext{e}{X-ray bolometric luminosity $L_{500}$ and its error
  in units of $10^{42}$\ erg\ s$^{-1}$.}  \tablefoottext{f}{Cluster
  mass $M_{500}$ and its error in units of $10^{13}$\ M$_\odot$.}
\tablefoottext{g}{Parameters derived from the cross-matching process
  of the cluster candidate sample and published cluster catalogues,
  see Sect. 3 for more detail.}  \tablefoottext{h}{The number of
  galaxies with spectroscopic redshifts consistent with the cluster
  redshift, when available, from SDSS-DR12 and their average. If there is no available value, we fill its space by -9999.}



\clearpage


\addtocounter{table}{-2}

\begin{table}
\caption{\label{tab:cand-cat} Cluster candidate sample (40 detections) with their X-ray positions, observation IDs, and fluxes as given in the 3XMM-DR5 catalogue.}
\end{table}

{\scriptsize 
\begin{longtable}{c c c c c c c c c c c c c c }
 \hline
 \hline
  \multicolumn{1}{c}{DETID\tablefootmark{a}} &
  \multicolumn{1}{c}{IAU Name\tablefootmark{a}} &
  \multicolumn{1}{c}{RA\tablefootmark{a}} &
  \multicolumn{1}{c}{Dec\tablefootmark{a}} &
  \multicolumn{1}{c}{OBSID\tablefootmark{a}} &
  \multicolumn{1}{c}{$F_{cat}$\tablefootmark{a,c}} &
  \multicolumn{1}{c}{$\pm eF_{cat}$} &
  \multicolumn{1}{c}{Nr. \tablefootmark{a}} &
  \multicolumn{1}{c}{class\tablefootmark{b}} &
  \multicolumn{1}{c}{redshift\tablefootmark{c}} \\

  \multicolumn{1}{c}{} &
  \multicolumn{1}{c}{} &
  \multicolumn{1}{c}{(deg)} &
  \multicolumn{1}{c}{(deg)} &
  \multicolumn{1}{c}{} &
  \multicolumn{1}{c}{} &
  \multicolumn{1}{c}{} &
  \multicolumn{1}{c}{DETIDs} &
  \multicolumn{1}{c}{(X-ray)} &
  \multicolumn{1}{c}{(expected)} \\

(1) &  (2)  &  (3)  & (4)  &   (5)   &  (6)  &   (7)  &   (8)  &  (9) & (10)    \\
\hline 
103057510010070 & 3XMM J000436.8+000148 &   1.15346 &   0.03028 & 0305751001 &   0.80 & 0.21 & 1 & 1 &    low \\
104037603010026 & 3XMM J001200.5+005233 &   3.00223 &   0.87606 & 0403760301 &   3.49 & 0.41 & 1 & 1 &   high \\
104037601010039 & 3XMM J001703.1-005906 &   4.26323 &  -0.98501 & 0403760101 &   0.19 & 0.26 & 1 & 2 &   high \\
104037601010034 & 3XMM J001725.4-005734 &   4.35605 &  -0.95966 & 0403760101 &   0.55 & 0.25 & 1 & 2 &    low \\
104037606010004 & 3XMM J001738.1-005150 &   4.40884 &  -0.86411 & 0403760601 &   5.55 & 2.45 & 1 & 2 &   high \\
104070301010074 & 3XMM J002302.7+000842 &   5.76153 &   0.14520 & 0407030101 &   0.10 & 0.19 & 1 & 2 &   high \\
104031601010027 & 3XMM J002905.7-001639 &   7.27381 &  -0.27767 & 0403160101 &   1.00 & 0.59 & 1 & 2 &   high \\
102036901010131 & 3XMM J003850.0+010401 &   9.70837 &   1.06706 & 0203690101 &   0.50 & 0.22 & 1 & 2 &   high \\
103035622010025 & 3XMM J004356.9+000743 &  10.98736 &   0.12869 & 0303562201 &   7.19 & 0.98 & 1 & 2 &   high \\
102007306010046 & 3XMM J021021.7-000721 &  32.59055 &  -0.12264 & 0200730601 &   1.24 & 0.31 & 1 & 1 &    low \\
106524006010090 & 3XMM J022731.7+002951 &  36.88241 &   0.49762 & 0652400601 &   0.75 & 0.57 & 1 & 2 &   high \\
106524006010077 & 3XMM J022734.8+004134 &  36.89501 &   0.69295 & 0652400601 &   0.26 & 0.45 & 1 & 2 &   high \\
106524006010067 & 3XMM J022740.0+003926 &  36.91706 &   0.65747 & 0652400601 &   0.30 & 0.13 & 1 & 1 &    low \\
106524006010056 & 3XMM J022742.2+004150 &  36.92589 &   0.69739 & 0652400601 &   0.76 & 0.96 & 1 & 1 &    low \\
106524006010094 & 3XMM J022756.2+003238 &  36.98430 &   0.54396 & 0652400601 &   1.10 & 0.59 & 1 & 1 &    low \\
106524006010070 & 3XMM J022834.3+004447 &  37.14330 &   0.74639 & 0652400601 &   1.56 & 0.27 & 1 & 1 &   high \\
106524007010092 & 3XMM J022931.9+004459 &  37.38322 &   0.74990 & 0652400701 &   0.32 & 0.43 & 1 & 1 &   high \\
106524007010093 & 3XMM J022937.1+003947 &  37.40486 &   0.66322 & 0652400701 &   0.19 & 0.14 & 1 & 1 &   high \\
106524008010104 & 3XMM J023052.1+005048 &  37.71730 &   0.84668 & 0652400801 &   0.56 & 0.25 & 1 & 2 &   high \\
100411701010109 & 3XMM J030213.0+000559 &  45.55438 &   0.09995 & 0041170101 &   0.26 & 0.05 & 1 & 1 &   high \\
100411701010111 & 3XMM J030237.8+000027 &  45.65756 &   0.00768 & 0041170101 &   0.17 & 0.05 & 1 & 1 &   high \\
101426101010025 & 3XMM J030631.9+000113 &  46.63330 &   0.02031 & 0142610101 &   0.68 & 0.09 & 1 & 2 &   high \\
102011201010105 & 3XMM J030642.2-001222 &  46.67596 &  -0.20612 & 0201120101 &   0.89 & 0.44 & 1 & 2 &   high \\
102011201010049 & 3XMM J030652.9-001121 &  46.72050 &  -0.18939 & 0201120101 &   0.54 & 0.10 & 1 & 1 &    low \\
101426101010088 & 3XMM J030653.7-000309 &  46.72399 &  -0.05272 & 0142610101 &   0.23 & 0.09 & 1 & 2 &    low \\
102032305010003 & 3XMM J035414.6+003703 &  58.56112 &   0.61771 & 0203230501 &   7.20 & 0.85 & 1 & 1 &    low \\
103048012010063 & 3XMM J213311.8-003609 & 323.29926 &  -0.60270 & 0304801201 &   0.62 & 0.87 & 1 & 2 &   high \\
100943101010077 & 3XMM J221715.0+001201 & 334.31142 &   0.20050 & 0094310101 &   0.37 & 0.05 & 2 & 1 &   high \\
100943101010032 & 3XMM J221744.1+001721 & 334.43404 &   0.28940 & 0094310101 &   1.14 & 0.07 & 2 & 1 &   high \\
100943101010089 & 3XMM J221810.4+001931 & 334.54269 &   0.32617 & 0094310101 &   0.57 & 0.10 & 2 & 1 &   high \\
100943102010074 & 3XMM J221830.2+001221 & 334.62554 &   0.20655 & 0094310201 &   0.81 & 0.12 & 2 & 2 &    low \\
106700202010014 & 3XMM J222116.2-010031 & 335.31774 &  -1.00882 & 0670020201 &   2.94 & 0.59 & 1 & 1 &    low \\
106524010010060 & 3XMM J232739.4+000650 & 351.91443 &   0.11413 & 0652401001 &   0.59 & 0.09 & 1 & 1 &   high \\
106524010010032 & 3XMM J232751.3+000202 & 351.96413 &   0.03412 & 0652401001 &   0.82 & 0.12 & 1 & 1 &   high \\
106524011010077 & 3XMM J232850.3+001356 & 352.20977 &   0.23222 & 0652401101 &   0.48 & 0.09 & 1 & 1 &    low \\
106524011010058 & 3XMM J232853.8+000540 & 352.22439 &   0.09466 & 0652401101 &   0.66 & 0.10 & 1 & 1 &    low \\
106524012010085 & 3XMM J233005.4+002113 & 352.52284 &   0.35363 & 0652401201 &   0.34 & 0.40 & 1 & 1 &    low \\
106524012010086 & 3XMM J233008.6+001528 & 352.53615 &   0.25797 & 0652401201 &   0.25 & 0.48 & 1 & 1 &   high \\
106524012010071 & 3XMM J233010.9+001853 & 352.54551 &   0.31499 & 0652401201 &   0.99 & 0.50 & 1 & 1 &   high \\
103031108010045 & 3XMM J235832.1-001523 & 359.63386 &  -0.25663 & 0303110801 &   0.41 & 0.30 & 1 & 2 &    low \\

\hline
\end{longtable}
}

\tablefoot{The cluster candidate catalogue is also available at CDS.}
\tablefoottext{a}{Parameters for X-ray cluster candidates extracted from the 3XMM-DR5 catalogue.} 
\tablefoottext{b}{A flag indicating the reliability of X-ray detection as class=1 (good candidate) and class=2 (weak candidate).}      
\tablefoottext{c}{A flag indicating the expected redshift of the candidate as low redshift ($z < 0.6$) and high redshift ($z > 0.6$).} 

\twocolumn


\end{document}